\documentclass[pra,aps,amssymb,twocolumn,noshowpacs]{revtex4}
\usepackage{color}
\usepackage{graphicx}
\usepackage{hyperref}
\usepackage{amsmath}
\usepackage{latexsym}
\usepackage{amssymb}
\usepackage{mathrsfs}
\usepackage{mathtools}
\usepackage{layout}
\usepackage{verbatim}
\usepackage{bm}
\usepackage{amsfonts,epsfig}

\usepackage{tikz}
\newcommand\circled[1]{\tikz[baseline=(char.base)]{
            \node[shape=circle,draw,inner sep=0.5pt] (char) {#1};}}
            
\begin{document}
\title{Annulled van der Waals interaction and fast Rydberg quantum gates }

\date{\today}
\author{Xiao-Feng Shi}
\affiliation{School of Physics, Georgia Institute of Technology, Atlanta, GA, 30332-0430, USA}

\author{T. A. B. Kennedy}
\affiliation{School of Physics, Georgia Institute of Technology, Atlanta, GA, 30332-0430, USA}

\begin{abstract}
  A pair of neutral atoms separated by several microns and prepared in identical s-states of large principal quantum number experience a van der Waals interaction. If microwave fields are used to generate a superposition of s-states with different principal quantum numbers, a null point may be found at which a specific superposition state experiences no van der Waals interaction. An application of this novel Rydberg state in a quantum controlled-Z gate is proposed, which takes advantage of GHz rate transitions to nearby Rydberg states. A gate operation time in the tens of nanoseconds is predicted.

\end{abstract}
\maketitle
\section{introduction}
The realization of a quantum computer requires scalable systems of quantum bits~(qubits) to perform fast gate operations ~\cite{Ladd2010}. Progress towards achieving robust qubits, characterized by speed and coherence, involves several different physical platforms, from the solid state systems of semiconductor quantum dots~\cite{Loss1998,Awschalom2013} and superconducting circuits~\cite{You2005,Chuang2013} to atomic ions~\cite{Cirac1995,Blatt2008} and neutral Rydberg atoms~\cite{PhysRevLett.85.2208,Saffman2010}.

 While solid state qubits may be manipulated in the nanosecond regime, they are more susceptible to environmental decoherence than atomic qubits~\cite{Bluhm2010,Awschalom2013,Xiang2013}. Atoms and ions have longer coherence times but correspondingly longer quantum gate times ~\cite{Monz2011,Ballance2015,Tan2015}. Superconducting qubits are promising both for large-scale quantum computations \cite{Rigetti2012,Barends2014}, and for hybrid systems with other more stable qubits~\cite{Xiang2013}.

Neutral alkali-metal atoms interact strongly at $\mu$m scale separation, by dipolar or van der Waals interactions (vdWI), when excited to Rydberg states of high principal quantum number $\mathsf{n}$. Such interactions have enabled the design of two-qubit quantum gates, based on phase-shift~\cite{PhysRevLett.85.2208} or Rydberg blockade~\cite{Isenhower2010,Wilk2010,Zhang2010,Maller2015,Jau2015}. The phase-shift gate can in principle operate on nanosecond  timescales, although this difficult challenge would require driving Rydberg states with an effective laser Rabi frequency in the GHz range.

The Rydberg blockade quantum gate time is limited by the effective Rabi frequency, $\Omega$, of the fields that rotate the qubit basis states~\cite{Saffman2016}. The blockade effect is caused by the vdWI shifting the two-atom Rydberg state by a characteristic frequency $\mathbb{B}.$ To avoid spurious laser excitation of blockaded states $\Omega \ll \mathbb{B},$ so that the two-qubit gate time is necessarily longer than $2\pi/\mathbb{B}$~\cite{Saffman2005}; for example, $\mathbb{B}/(2\pi)$ is about $40$~MHz when the atomic separation $L \sim 10\mu$m and $\mathsf{n} =  100$~\cite{Saffman2010}. To achieve sub-microsecond quantum gates, Refs.~\cite{Muller2011,Goerz2014} suggested optimal control of laser pulses with very large peak Rabi frequencies, in order to increase speed without increasing errors.

\begin{figure}
\includegraphics[width=3.3in]
{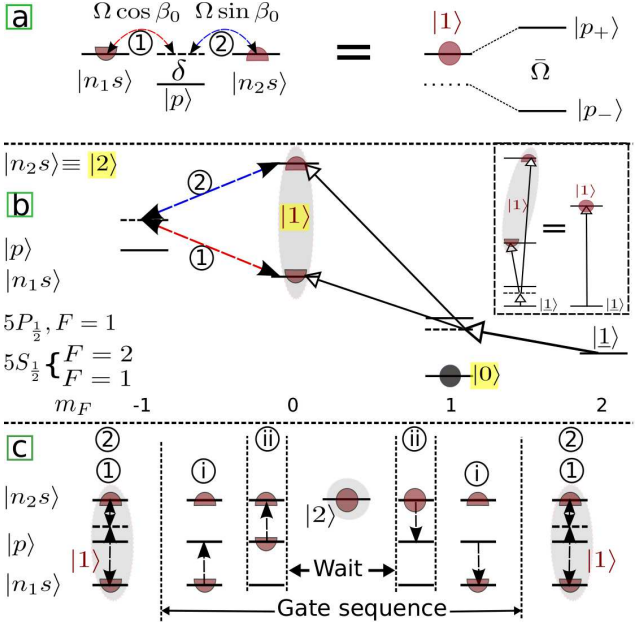}
 \caption{ Superposition states are illustrated by half discs and eigenstates by discs. (a) Dressed state picture for three Rydberg states coupled by two microwave fields. The state $|1\rangle$ is a dark state, and $|p_\pm\rangle$ are defined below Eq.~(\ref{dressing}) (b) A schematic of the atomic levels of $^{87}$Rb used in preparing the state $|1\rangle$ along with the laser and microwave fields. Inset: illustrates that the Y configuration of lasers produces an effective two state system. (c) State evolution of one atom during the $C_{\text{Z}}$ gate protocol involving two pairs of microwave pulses, denoted by i and ii, separated by a wait period.  \label{fig001} }
\end{figure}

Here we propose a new type of Rydberg quantum gate that operates on a timescale of tens of nanoseconds, by exploiting the properties of a special Rydberg state, a superposition of s-states $|\mathsf{n}_{1}s\rangle$ and $|\mathsf{n}_{2}s\rangle,$ with different principal quantum numbers. The state, denoted by $|1 \rangle,$ is designed to experience neither Rydberg blockade shift nor vdWI-induced decay, and is formed by coupling the s-states to a nearby p-orbital with a pair of microwave fields in an Autler-Townes configuration, see Fig.~\ref{fig001}(a). Under these conditions $|1\rangle$ is a dark-state of the microwave field~(it contains no $|p\rangle$-state admixture). We note that the mixing angle $\beta$ in the superposition state, $\sin \beta |\mathsf{n}_{1}s\rangle -\cos \beta |\mathsf{n}_{2}s\rangle$ may be tuned by varying the relative strength of the two microwave field Rabi frequencies in Fig.~\ref{fig001}(a). By tuning $\beta$ in this way it is possible to access a point $\beta = \beta_0$ where the blockade shift of the product dark state $|1,1 \rangle$ vanishes, provided an appropriate choice of principal quantum numbers has been made. Hence we may write $|1 \rangle$ explicitly in the form
\begin{equation} \label{darkstate}
|1 \rangle =\sin \beta_0 |\mathsf{n}_{1}s\rangle -\cos \beta_0 |\mathsf{n}_{2}s\rangle.
 \end{equation}
 An unusual feature of our gate protocol is that $|1\rangle,$ a radiatively metastable state, is employed as a qubit basis state, while the other basis state, denoted $|0\rangle$, is a ground hyperfine state.
 Since the state $|1\rangle$ can be microwave coupled to an auxiliary Rydberg eigenstate, denoted $|2 \rangle$, with a GHz-scale Rabi frequency much larger than the blockade shift $\mathbb{B}$ of the state $|2 ,2 \rangle$, the phase accumulated in this state due to $\mathbb{B}$ may be used to implement a controlled-Z~($C_{\text{Z}}$) gate. In this case $2 \pi / \mathbb{B}$ determines the quantum gate operation time, and as shown below, leads to $C_{\text{Z}}$ gate operation times of tens of ns. Accessing this regime with high fidelity requires high-frequency microwave sources in the range $100$ to $200$~GHz, such as those employed in studies of electron paramagnetic resonance~\cite{Nafradi2008,Blok2004,Takahashi2008,Cho2014,Siaw2016}.
Stabilization of the qubit to radiative decay may be performed by mapping $|1 \rangle$ to an electronic ground state $| \underline{1} \rangle$ by a laser pulse.

This article is organized as follows. In Sec.~\ref{sec002}, we describe how a superposition state of different two-atom Rydberg states can have zero vdWI. After introducing our $C_{\text{Z}}$ gate protocol in Sec.~\ref{sec003}, we present how the superposed state approximates an eigenstate in the presence of vdWI during the gate protocol in Sec.~\ref{sec004}. Section~\ref{sec005} details error estimates about the gate fidelity, Sec.~\ref{sec006} presents a scheme of stabilizing the gate, and Sec.~\ref{sec007} gives a summary. Additional details of the theory are given in the Appendixes.

\begin{figure}
\includegraphics[width=3.3in]
{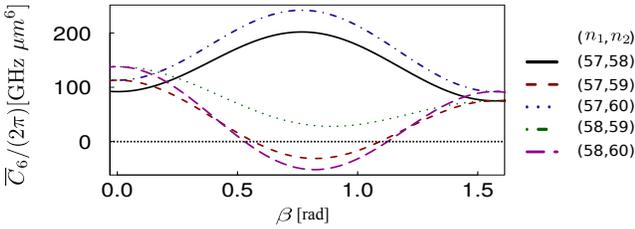}
 \caption{ The vdWI coefficient $\overline{C}_6$ (see text) as a function of $\beta$ for different pairs of principal quantum numbers $(\mathsf{n}_1,\mathsf{n}_2)$. The Rydberg states have a common magnetic quantum number $m_J=-1/2$. \label{fig002}}
\end{figure}

\section{Non-interacting two-atom Rydberg dark state}\label{sec002}
We wish to prepare each of a pair of atoms, the control and target of a two qubit quantum gate, in the state $|1\rangle$, Eq.~(\ref{darkstate}). Introducing the orthogonal superposition states~[see Eq.~(\ref{jiajian1}) for the derivation],
\begin{eqnarray}
  \begin{array}{c}
   |+ \beta\rangle=\cos\beta |\mathsf{n}_{1}s  \rangle +\sin\beta |\mathsf{n}_{2}s \rangle,\\
   |- \beta\rangle=\sin\beta |\mathsf{n}_{1}s  \rangle-\cos\beta |\mathsf{n}_{2}s \rangle,
   \end{array}\label{r1definition}
 \end{eqnarray}
 we investigate the Rydberg blockade shift of the product state $|- \beta ,- \beta \rangle$ as a function of $\beta,$ and show that an angle $\beta_0$ can be chosen so that $|1 ,1 \rangle \equiv |- \beta_0 ,- \beta_0 \rangle$ experiences zero blockade shift.

When the distance $L$ between the atoms is sufficiently large, the electric dipole-dipole interaction will only couple states of comparable energy and this determines the vdWI regime. In the presence of the microwave fields introduced below, we may transform to a rotating frame in which the ordered set of states $(|\mathsf{n}_{1}s\mathsf{n}_{1}s \rangle,|\mathsf{n}_{1}s \mathsf{n}_{2}s \rangle,|\mathsf{n}_{2}s \mathsf{n}_{1}s \rangle,|\mathsf{n}_{2}s \mathsf{n}_{2}s \rangle)$ is degenerate in the absence of vdWI. We may then describe the atomic interactions in the presence of the fields by the dressed vdWI operator $\hat {H}_{v} = h^{(0)}_v + h^{(1)}_v,$ where
$h^{(0)}_v = \mbox{diag}(a,b,b,d)/L^6$ is diagonal in the ordered basis and $h^{(1)}_v = c\left[|\mathsf{n}_{1}s \mathsf{n}_{2}s \rangle\langle \mathsf{n}_{2}s \mathsf{n}_{1}s |+\text{h.c.}\right]/L^6,$ is an interaction between degenerate two-atom states. The operator given assumes the absence of quasi-degenerate levels $(\mathsf{n}_{1}'p_{1(3)/2},\mathsf{n}_{2}'p_{1(3)/2})$.

 In terms of the superposition states of Eq.~(\ref{r1definition}), $\hat H_{v}$ may be written~[see Appendix~\ref{app_B} for details]
 \begin{eqnarray}
   \hat{H}_{v}&=&\frac{1}{L^6}
   \prod_{k=1}^4 \sum_{v_k = \pm \beta}
   C_6(v_1v_2v_3v_4)|v_1,v_2\rangle\langle v_3,v_4|, \label{eq03}
 \end{eqnarray}
where $ \langle - \beta, - \beta| \hat H_{v}| - \beta, - \beta \rangle = \overline{C}_6(\beta)/L^6 $, and
  \begin{eqnarray}
\overline{C}_6(\beta)= a\sin^4\beta+ d\cos^4\beta+2(b+c)\sin^2\beta\cos^2\beta.\nonumber
  \end{eqnarray}
  The key point is that while the coefficients $a$ and $d$ are positive for $s$-orbital Rydberg states, $b$ may be negative~\cite{PhysRevLett.115.013001}. To realize $\overline{C}_6=0$, it is necessary to select Rydberg levels with a sufficiently large and negative $b,$ and a suitable angle $\beta = \beta_0$, in order to cancel the positive contributions from $a$ and $d$.
Numerical results for $\overline{C}_6$ for five different pairs of rubidium Rydberg levels $(\mathsf{n}_{1},\mathsf{n}_{2})$ are shown in Fig.~\ref{fig002}. In two of these cases it is possible to find a $\beta = \beta_0$ for which $\overline{C}_6 = 0.$ For example, $\overline{C}_6=0$ for $(\mathsf{n}_1,\mathsf{n}_2,\beta_0)=(57,59,0.566)$ corresponding to $(a,b,c,d)/(2 \pi) = (75,-149,-6,113)$~GHz~$\mu \text{m}^6$~\cite{Walker2008,Shi2014}. This example will be used in our analysis of the $C_Z$ gate protocol.

Figure~\ref{fig001}(b) illustrates how to initialize the state $|1\rangle$ starting from the atomic ground state $|\underline{1}\rangle=|5S_{1/2}, F=2,m_F=2\rangle$, in the presence of the microwave dressing and vdWI. Here the microwave dressing with the two fields \circled{1} and \circled{2} in Fig.~\ref{fig001}, and the Hamiltonian $\hat{H}_M$, as shown later on in Eq.~(\ref{dressing}), can protect $|1,1\rangle$ against vdWI-induced decay~[see its numerical test in Appendix~\ref{app_C}; Also see Eq.~(\ref{eqA10})]. Two-photon excitations with effective Rabi frequencies $\kappa_{1}$ and $\kappa_{2}$ couple the ground state $|\underline{1}\rangle$ to $|\mathsf{n}_1s\rangle $ and $|\mathsf{n}_2s\rangle $, respectively, via a low energy p-state. The full Hamiltonian during the initialization reads
\begin{eqnarray}
 \hat{H}_{0} &=&\hat{H}_{v} +\sum_{i=\text{c,t}} \left( \hat{H}_{M} + \sum_{j=1}^2(\kappa_{j}|\mathsf{n}_js\rangle\langle \underline{1}|+\text{h.c.})/2\right)_i,\label{iniH}
\end{eqnarray}
where $i$ labels the control~c and target~t atoms, respectively. When $\kappa_{2}=-\cot\beta_0\kappa_{1}$, it is readily shown that only $|1\rangle$ is coupled to $|\underline{1}\rangle$, while $|p_\pm\rangle$ are decoupled. Moreover, as the two-atom state $|1,1 \rangle$ does not suffer a blockade or ac Stark shift, it will be resonantly excited when the atoms are initially prepared in the state $|\underline{1},\underline{1}\rangle$.

It is essential to be able to tune $\beta$ to the value $\beta_0$ that determines the two-particle dark state $|1,1\rangle$ in the presence of the microwave dressing fields. That this condition is satisfied can be recognized by observing Rabi oscillation cycles as follows~\cite{Gaetan2009,Urban2009}. Using a resonant laser coupling between the two atom ground state $|\underline{1},\underline{1}\rangle$ and Rydberg state $|- \beta, -\beta\rangle$, and a small enough laser Rabi frequency, observation of a complete Rabi oscillation indicates that $\beta = \beta_0$. If an incomplete Rabi oscillation is observed, that is the ground state population does not reach zero, then change $\beta$ in order to minimize the ground state population and continue until a complete Rabi oscillation is observed. Details about how to tune the mixing angle to $\beta_0$ can be found in Appendix~\ref{app_D}.

Assuming that $\beta_0$ has been found in this way, the required condition $\kappa_{2}=-\cot\beta_0\kappa_{1}$ on the laser Rabi frequencies may be identified by observing an undepleted ground state $|\underline{1}\rangle$ when the lasers are resonantly tuned to the states $|p_\pm\rangle$ of one of the atoms~(See Appendix~\ref{app_E} for more details).

\section{ $C_{\text{Z}}$ gate protocol}\label{sec003}
A $C_{\text{Z}}$ gate is the state transformation: $\alpha_1 |0,0\rangle + \alpha_2 |0,1\rangle + \alpha_3 |1,0\rangle+ \alpha_4 |1,1\rangle \mapsto \alpha_1 |0,0\rangle + \alpha_2 |0,1\rangle + \alpha_3 |1,0\rangle - \alpha_4 |1,1\rangle$ ~\cite{Nielsen2000}.
 In order to implement the $C_{\text{Z}}$ gate, a microwave source induces a fast transition from $|1\rangle$ to another Rydberg state $|2\rangle$~\cite{Ryabtsev2003,Afrousheh2004,Ryabtsev2010,Anderson2014,Arakelyan2016}, such that $|2, 2 \rangle$ experiences a large blockade shift $\mathbb{B}$ due to the vdWI. Accessing the state $|2,2 \rangle$ is subject to an error due to its Rydberg blockade shift, but this error is suppressed with moderately large microwave field strengths. After generating state $|2,2\rangle$, a waiting period of duration $T=\varphi/\mathbb{B}$ produces the state transformation $|2,2\rangle\mapsto e^{-i\varphi}|2,2\rangle$. The accumulated phase is transferred to $|1,1\rangle$ when $|2\rangle$ is excited back to $|1\rangle$, realizing a $C_{\text{Z}}$ gate if $\varphi=\pi$.

  A five stage $C_{\text{Z}}$ gate protocol involving a sequence of four microwave pulses, the first and second pair separated by a waiting period is shown in Fig.~\ref{fig001}(c). The microwave dressing field is switched off at the beginning of this protocol and is switched back on at its completion. If we choose $|2\rangle\sim|\mathsf{n}_2s\rangle$, then $|22\rangle$ has a vdWI coefficient $C_6/(2 \pi)=113$~GHz~$\mu \text{m}^6$. For a two-atom separation of $L=4.4\mu m$, then $\mathbb{B}/(2\pi)=16$~MHz. The transition between s-states $|1\rangle$ and $|2\rangle$ involves a microwave coupling to an intermediate p-orbital which we choose to be the same state $|p \rangle$ used in the preparation of $|1\rangle$. Specifically we take $|p\rangle\equiv |54p_{3/2},m_J=-3/2\rangle$. This choice avoids leakage of population in the transitions between $|1\rangle$ and $|2 \rangle$ and in the microwave dressing process. The transition frequencies between $|\mathsf{n}_{1(2)}s\rangle$ and $|p\rangle$ are in the range $100- 200$~GHz.

\begin{figure}
\includegraphics[width=3.3in]
{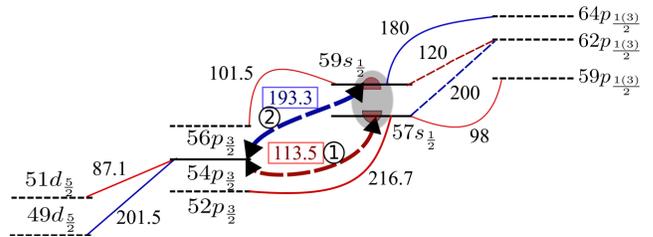}
\caption{Atomic levels involved in the $C_{\text{Z}}$ gate protocol and loss channels. The two thick dashed curves denote the two chosen transitions, the two thin dashed curves denote a resonant two-photon process, while all other solid curves denote one-photon loss channels, with corresponding transition frequencies given~(in units of GHz).  \label{fig003}}
\end{figure}
   During the gate sequence, both fields \circled{1} and \circled{2} are switched off. The gate sequence is
\begin{eqnarray}
 |\mathsf{n}_1s\rangle\xrightarrow[\Omega_1]{\pi} |p\rangle \xrightarrow[\Omega_2]{2\beta_0} |\mathsf{n}_2s\rangle \xrightarrow[\text{wait}]{T} |\mathsf{n}_2s\rangle \xrightarrow[\Omega_2]{2\beta_0} |p\rangle \xrightarrow[\Omega_1]{\pi} |\mathsf{n}_1s\rangle,\label{3levle}
\end{eqnarray}
where $\Omega_j/\Omega = \cos\beta_0~(\sin\beta_0)$ for $j=1~(2)$.
Here the first and last transformations involve $\pi$ pulses of microwave field \circled{i}, while the other two transformations involve microwave field \circled{ii} and have pulse area $2\beta_0$. The full Hamiltonian for the first and second transformations above is $\hat{H}_{v}+\hat{H}_{j} $, for $j=1,2$ respectively, where
\begin{eqnarray}
\hat{H}_{j} &=&\Omega_j\sum_{i=\text{c,t}}(|\mathsf{n}_js\rangle\langle p|+\text{h.c.})/2]_i. \label{hamil_j}
  \end{eqnarray}
As is easily shown in Appendix~\ref{App_F}, at the end of the first two transformations each atom is in the state $|2\rangle$, and after the waiting period, unitary evolution generated by $\hat{H}_2$ and $\hat{H}_1$ consecutively, completes the gate process. At the end of the fourth pulse, microwave dressing fields \circled{1} and \circled{2} in Fig.~\ref{fig001} are switched back on, so that the system Hamiltonian returns to Eq.~(\ref{iniH}) with $\kappa_1=\kappa_2=0$.

\section{Suppression of vdWI-induced decay}\label{sec004}
We now show that the state $|1,1\rangle,$ defined by the microwave dressing fields, is protected from vdWI-induced decay to other two-particle states. The Rydberg states $|\mathsf{n}_1 s\rangle$ and $|\mathsf{n}_2 s\rangle$ are coupled to a common p-state, denoted $|p\rangle$, by two microwave fields, as shown in Fig.~\ref{fig001}. The microwave field ac Stark shifts nearby two-atom states out of resonance, without perturbing the state $|1,1\rangle$ in any way.
 The two-photon resonant coupling of the $|\mathsf{n}_1 s \rangle$ and $|\mathsf{n}_2 s \rangle$ states via $|p\rangle,$ with respective Rabi frequencies $\Omega\cos\beta_0$ and $\Omega\sin\beta_0$, is one-photon detuned by $\delta$. Then, in the rotating frame introduced above, the interaction with the microwave fields is given by~(see Appendix~\ref{App_A})
 \begin{eqnarray}
\hat{H}_{M}&=&\text{diag}(\bar{\Omega}+\delta,0,-\bar{\Omega}+\delta)/2, \label{dressing}
 \end{eqnarray}
where
 $\hat{H}_{M}$ is written in the ordered basis $\{|p_+\rangle, |1\rangle,  |p_-\rangle\}$, with $|p_{+}\rangle =  \sin \gamma |+ \beta_0 \rangle + \cos \gamma|p\rangle$, $|p_{-}\rangle =  \cos \gamma |+ \beta_0 \rangle - \sin \gamma|p\rangle$. The generalized Rabi frequency $\bar \Omega = \sqrt{\Omega^2+\delta^2}$ and $\tan 2 \gamma = \Omega/\delta.$ The microwave excitation fields are labeled as \circled{1} and \circled{2} in Fig.~\ref{fig001}. Note that the state $|1\rangle$ is unperturbed by the microwave dressing, i.e., it is a dark state. The choice of $\Omega$ and $\delta$ in Eq.~(\ref{dressing}) strongly influences the vdWI between $|1,1\rangle$ and the other three basis states in Eq.~(\ref{eq03}). This is because only the state $|1,1\rangle$ in Eq.~(\ref{eq03}) remains a good eigenstate of the microwave field Hamiltonian in Eq.~(\ref{dressing}), while the other three states $|+\beta_0,+\beta_0\rangle,|-\beta_0,+\beta_0\rangle,|+\beta_0,-\beta_0\rangle$ split and shift into the set of eight eigenstates $|-p_\pm\rangle,|p_\pm-\rangle,|p_\pm p_ \pm\rangle,|p_\pm  p_\mp\rangle$, which are separated in energy from $|1,1\rangle$ by energy gaps $\Delta E \in \{\delta, \delta \pm \bar{\Omega}, (\delta \pm \bar{\Omega})/2 \}$ that we want to be large compared with any dipole matrix element in the two-atom system. In this case, the dressed vdWI coupling from $|1,1\rangle$ to these other states is suppressed.

We now address the residual effects of the microwave field on the dressed vdWI~(see Appendix~\ref{app_B}). We note that the state $|p\rangle$ has negligible contribution to the dressed vdWI of $|1,1\rangle$. The dipole matrix elements coupling $|p\rangle$ with $|\mathsf{n}_{j}\rangle$ ($j =1,2$), respectively, will be suppressed if we make the difference in their principal quantum numbers large enough; this will be satisfied in the example presented later. Moreover, as the $|1,1 \rangle$ state is unshifted by the microwave coupling, the dressed vdWI is negligibly affected by the $|p\rangle$ channel. Of course the channels, labeled $k$, that dominate the dressed atom vdWI of $|1,1\rangle$ are also modified by off-resonant ac-Stark shifts. Suppose that these shifts change the energy defects relative to the $|1,1\rangle$ state by $\Delta_k \mapsto \Delta_k + \delta_k$, producing a correction to the channel-k vdWI contribution of $O(\delta_k/ \Delta_k)$ \cite{Walker2008}. For the examples of relevance here this change is rather small: $\delta_{k}/\Delta_{k}$ is about $5\%$ for the strongest two vdWI channels, and decreases rapidly for weaker channel contributions. To leading order, therefore, the off-resonant microwave field ac Stark shifts do not significantly modify the dressed vdWI for $|1,1\rangle$ discussed above.

\section{Error estimates}\label{sec005}
The total error for the gate may be written as~(see Appendix~\ref{App_G})
$$
E_0 = E_{\tau} + E_1 + E_2 + E_B,
$$
where $E_{\tau}$ is the error due to radiative decay of the Rydberg states, $E_1$ and $E_2$ are errors due to one-photon and two-photon transitions~(see Fig.~\ref{fig003}) causing leakage of population from the gate, and $E_B$ is an error due to the blockade shift of $|2,2\rangle$. The change of the mixing angle $\beta$ induced by different decay rates of the two Rydberg states $|\mathsf{n}_1s\rangle$ and $|\mathsf{n}_2s\rangle$, the vdWI and dipole exchange processes between $s$ and $p$ states, and the vdWI of two atoms in $p$ states, can be neglected~(see Appendixes~\ref{App_H} and~\ref{App_I}). Extra details about these estimates can be found in Appendixes~\ref{App_J} and~\ref{App_K}.

We note that $E_1$ and $E_2$ depend on the detuning for each leaking channel, which is a function of $\delta$. Hence $E_0$ is a function of $\Omega$, $\delta$ and $\mathbb{B}$ (or equivalently the atomic separation $L$). Fixing $\mathbb{B}$ we may minimize $E_0$ with respect to the pair $(\Omega,\delta)$. For example, for $\mathbb{B}/(2\pi)=16$~MHz, we find $(\Omega,\delta)=2 \pi (0.65,0.26)$~GHz giving $T_{\text{g}}\approx37$~ns and $E_0^{\text{min}}\approx4.6\times10^{-3}$. The value of $\Omega$ here corresponds to a microwave field strength of $440~(740)$~V/m for both \circled{1} (\circled{2}), and \circled{i} (\circled{ii}). While a strong microwave field may in principle lead to ionization of Rydberg states, a detailed analysis in Appendix~\ref{App_J} shows that ionization is negligible here. As reported in, for example, Ref.~\cite{Takahashi2008}, the required fast switching of microwave fields appears feasible. 

\begin{figure}
\includegraphics[width=3.3in]
{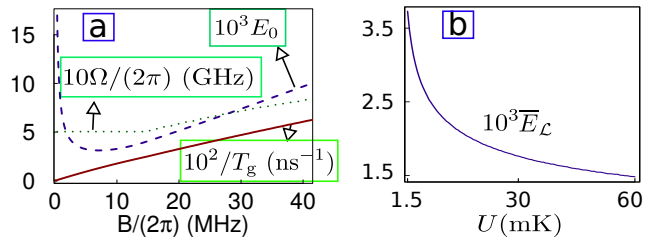}
\caption{(a) Performance of a $C_{\text{Z}}$ gate realized as a function of $\mathbb{B}$ when an optimal $\Omega$ is used for each $\mathbb{B}$. $T_{\text{g}}$: gate time, $E_0$: fidelity error~(Excluding $\overline{E}_{\mathcal{L}}$). Here $\delta/(2\pi)=0.26$~GHz. (b) Contribution~(scaled by $10^3$) to the gate fidelity error due to fluctuation of atomic positions as a function of the trap depth when $L=4.4\mu m$.  \label{fig004}}
\end{figure}

In Fig.~\ref{fig004}(a) we plot the resulting gate operation time $T_{\text{g}}$, optimized Rabi frequency and gate error as a function of $\mathbb{B}$. The minimum error shown, $3.1\times 10^{-3},$ is close to the gate error limit, $2 \times 10^{-3},$ of a conventional Rydberg $C_{\text{Z}}$ gate~\cite{Zhang2012,Keating2015}, while the gate operation time of around 50 ns, compares favorably to a gate time of several microseconds of a conventional gate. In the proposed gate, one photon transitions which leak population during the gate operation, limit the speed of the protocol and as a consequence radiative decay bounds the achievable error. 

In contrast to the protocols in~\cite{PhysRevLett.85.2208,Muller2009,Beterov2016}, the proposed gate is sensitive to fluctuations in atomic separation $L$ (that determines $\mathbb{B}$) due to atomic motion within the trapping potential. Ideally this could be minimized by cooling the qubits to their motional ground states as was done in~\cite{Kaufman2012}. With the trapping geometry of Refs.~\cite{Kaufman2012,Lester2015}, we have estimated the additional contribution $\overline{E}_{\mathcal{L}}$ to the gate error shown in Fig.~\ref{fig004}(b), where $U$ is the trap depth~(see Appendix~\ref{App_L}). We find that $\overline{E}_{\mathcal{L}}$ drops from $3.0\times10^{-3}$ to $1.5\times10^{-3}$ when $U$ increases from $3.5$~mK, as realized in~\cite{Lester2015}, to $60$~mK. Since attaining $U$ of order a few times $10$~mK is feasible for an optical trap~\cite{Saffman2016}, atomic separation errors may be suppressed. 

\section{Stabilized gates}\label{sec006}
The gate protocol involves the stable/metastable qubit states $\{|0\rangle,|1\rangle\}$. The qubit can be stabilized by means of a laser $\pi$ pulse which enforces the transition $|1\rangle \mapsto |\underline{1}\rangle,$ upon completion of the gate protocol. In this case, a stable qubit is defined by the pair $\{|0\rangle,|\underline1\rangle\}$. The $C_{\text{Z}}$ gate operation can thus be described as a three stage process: (I) initialization by laser $\pi$ pulse excitation $|\underline1\rangle \mapsto|1\rangle$ described by Eq.~(\ref{iniH}), (II) the gate sequence of Eq.~(\ref{3levle}), and (III) the qubit stabilizing $\pi$ pulse identical to that in step (I), mapping $|1\rangle \mapsto|\underline{1}\rangle$. The microwave fields \circled{1} and \circled{2} are switched on, except during step (II). The microwave and laser fields are far detuned from any excitation of the state $|0\rangle$. The two-photon laser Rabi frequency $\kappa\equiv \kappa_1\sin\beta_0-\kappa_2\cos\beta_0$ for steps (I) and (III) is bounded by the ground level $^{87}$Rb hyperfine splitting of $6.8$~GHz; for example $\kappa/(2\pi)=20$~MHz adds 50 ns, the gate initialization plus stabilization time, and increases the gate error by $2.8 \times 10^{-4}$ as a result of radiative decay of the Rydberg states.

\section{Conclusion}\label{sec007}
In conclusion, we have identified a special Rydberg atom superposition state, a dark state of applied microwave fields, and shown that the vdWI between two atoms each prepared in this state can be annulled by tuning the relative strength of the fields. We have presented a protocol in which the dark state serves as a qubit basis state for a quantum $C_Z$ gate. The gate operation cycle is predicted to be some tens of ns, thanks to GHz frequency transitions from the dark state to nearby Rydberg levels.

\section*{ACKNOWLEDGMENTS}
We acknowledge support from AFOSR MURI on Multifunctional light-matter interfaces based on neutral atoms and solids and thank M.S. Chapman and B. Hebbe Madhusdhana for discussions.

\appendix{}

\section{Derivation of $\hat{H}_{\text{M}}$}\label{App_A}
In this appendix and the following ones we will study the Hamiltonian of the microwave fields, dressed vdWI, interactions of states involving $|p\rangle$, state evolution during the gate sequence, methods for accessing the angle $\beta_0$, numerical verification of the dark state $|1,1\rangle$, ionization induced by microwave fields, amplitude fluctuations of the dressing microwave fields, and position fluctuation of the atoms.

The Hamiltonian of the system is $\hat{H}$, where
\begin{eqnarray}
  \hat{H}&=&\hat{H}_{\text{dd}} +\hat{H}_{\text{atom}} + \sum_{i=\text{c,t}}[\hat{H}_{\text{M}} + \hat{H}_{i} ] + \hat{H}_{k} ,\nonumber\\
 \hat{H}_{i}&=& \sum_{j=1}^2(\kappa_{j}|\mathsf{n}_js\rangle\langle \underline{1}|+\text{h.c.})/2,\nonumber\\
  \hat{H}_{k}&=& \Omega_k\sum_{i=\text{c,t}}[(|\mathsf{n}_js\rangle\langle p|+\text{h.c.})/2]_i,\nonumber
\end{eqnarray}
where $\hat{H}_{\text{dd}}$ is the dipole-dipole interaction, $\hat{H}_{\text{atom}}$ is the atomic energy term, while $\hat{H}_{\text{M}}$ of the microwave fields \circled{1} and \circled{2} are present only before and after the gate sequence, and $\hat{H}_{i}$ is from the optical lasers for initializing $|1\rangle$. Finally, $\hat{H}_{k}$ accounts for the four-pulse sequence of the microwave fields \circled{i} and \circled{ii}. The microwave fields \circled{1} (\circled{2}) and \circled{i} (\circled{ii}) differ only by a wavelength difference, that is, the detuning $\delta$. 

For the sake of convenience, we ignore the frequently appeared factor of $2\pi$ for frequencies.

We use an operator
\begin{eqnarray}
\hat{R}_0 &=& \sum_{j= n_1s,n_2s} E_j|j\rangle\langle j| + (E_p-\delta)|p\rangle\langle p|, \label{rotate0}
\end{eqnarray}
to transform the Hamiltonian into a rotating frame,
\begin{eqnarray}
\hat H_{\text{M}}+\hat{H}_{\text{atom}}
&\rightarrow&\hat  e^{i\hat{R}_0t }(\hat H_{\text{M}}+\hat{H}_{\text{atom}}) e^{-i\hat{R}_0t }- \hat{R}_0.\nonumber
\end{eqnarray}
After electric dipole and rotating-wave approximations and for a coupling scheme shown in Fig.~\ref{fig001}(a), the Hamiltonian above becomes
\begin{eqnarray}
  \hat{H}_{\text{M}}&=&\left(\begin{array}{ccc}
    0 & \Omega\cos\beta& 0\\
    \Omega\cos\beta& 2\delta& \Omega\sin\beta\\
    0 & \Omega\sin\beta &0
    \end{array} \right)/2, \label{HM001}
 \end{eqnarray}
written in the ordered basis of $\{|\mathsf{n}_{1}s  \rangle , |p\rangle,|\mathsf{n}_{2}s \rangle \}$. Here we do not include the part of $\hat{H}_{\text{atom}}$ that does not change under the rotating frame transformation. The diagonalization of the Hamiltonian above gives Eq.~(\ref{dressing}), i.e.,
\begin{eqnarray}
  \hat{H}_{\text{M}}&=&\left(\begin{array}{ccc}
   \Omega_{\text{p}}  +\delta& 0 &0 \\
   0&0 &0\\
    0 & 0&- \Omega_{\text{p}}+\delta
  \end{array} \right)/2, ~\Omega_{\text{p}}=\sqrt{\Omega^2+\delta^2},\nonumber\\
  \label{eqA4}
 \end{eqnarray}
written in the ordered basis of $\{|p_+\rangle, |-\rangle,  |p_-\rangle \}$, where $\Omega_{\text{p}}$ is the parameter $\bar{\Omega}$ of the main text. Here
\begin{eqnarray}
   |p_+\rangle&=&\sin\gamma |+ \rangle +\cos\gamma |p \rangle,\nonumber\\
   |p_-\rangle&=&\cos\gamma |+ \rangle -\sin\gamma |p \rangle,\nonumber\\
   \sin\gamma &=& \Omega/N_p, \cos\gamma = (\Omega_{\text{p}}+\delta)/N_p,\nonumber\\
   N_p&= & \sqrt{\Omega^2+(\Omega_{\text{p}}+\delta)^2},
   \label{pppm}
\end{eqnarray}
and
\begin{eqnarray}
   |+\rangle&=&\cos\beta |\mathsf{n}_{1}s  \rangle +\sin\beta |\mathsf{n}_{2}s \rangle,\nonumber\\
   |-\rangle&=&\sin\beta |\mathsf{n}_{1}s  \rangle-\cos\beta |\mathsf{n}_{2}s \rangle\equiv |1\rangle.\label{jiajian1}
\end{eqnarray}

The eigenvalue of $\hat{H}_{\text{M}}$ for the target state is zero
\begin{eqnarray}
  |--\rangle &:& 0,\label{eqA9}
\end{eqnarray}
while those of the other eight states have the following eigenvalues of $\hat{H}_{\text{M}}$,
\begin{eqnarray}
  |-p_+\rangle, |p_+-\rangle &:&(\Omega_{\text{p}}+\delta )/2 =480\text{MHz} ,\nonumber\\
  |-p_-\rangle, |p_--\rangle &:&(-\Omega_{\text{p}}+\delta )/2=-220\text{MHz}  ,\nonumber\\
  |p_+p_+\rangle &:&\Omega_{\text{p}}+\delta=960\text{MHz} ,\nonumber\\
  |p_-p_-\rangle &:&-\Omega_{\text{p}}+\delta =-440\text{MHz},\nonumber\\
  |p_+p_-\rangle ,|p_-p_+\rangle&:&\delta=260\text{MHz} , \label{eqA10}
\end{eqnarray}
where the numerical values above are from a typical example of the main text, where $(\delta,\Omega)=(0.26,0.65)$~GHz. The choice of $\delta$ should avoid unwanted dipole coupling with nearby levels as indicated by the thin solid curves in Fig.~\ref{fig003}. Since the energy difference between $|--\rangle$ and any of the states in Eq.~(\ref{eqA10}) is much larger than any dipole coupling of the two atoms with the distances $L$'s chosen in the main text, only direct dipole interaction can couple $|--\rangle$ to other states. The direct coupling rates are given in Eq.~(\ref{strongDiple}), which are, however, three orders smaller than the energy gaps, resulting of $|--\rangle$ being an eigenstate.

\section{VdWI in the presence of strong microwave fields}\label{app_B}
 Before and after the gate sequence, the Hamiltonian is
\begin{eqnarray}
  \hat{H}&=&\hat{H}_{\text{dd}} +\hat{H}_{\text{atom}} + \sum_{i=\text{c,t}}[\hat{H}_{\text{M}} + \hat{H}_{i} ] ,\nonumber
\end{eqnarray}
where $\hat{H}_{i}$ is weak compared to $\hat{H}_{\text{M}}$, while $\hat{H}_{\text{M}}$ is strong. In this case, the energy difference between two dipole-coupled states are effectively changed, and may induce new features for the vdWI. We will explain below why this does not alter our analysis about the spectral isolation of the state $|--\rangle$. During the gate sequence, there is also some energy shift whose magnitude is comparable to those appeared below. With these shifts, the vdWI for relevant states are still much smaller than the microwave field Rabi frequencies. Thus we still ignore these vdWI during the gate sequence.

Notice that the atomic separation $L=4.4\mu $m will be quoted frequently, for the sake of concreteness, and the following analysis takes the example of the main text where $\beta=0.566$ and $(\delta,\Omega)=(0.26,0.65)$~GHz. Changing $\Omega$ and $L$ slightly for achieving different vdWI of the state $|22\rangle$ as in Fig.~\ref{fig004} of the main text does not alter our conclusions.
\subsection{Polarizability of Rydberg atoms}
The first type of energy shift is due to the polarizability $\mathscr{P}_i$ of the electron in a Rydberg atom~\cite{Maller2015,Topcu2013} in the presence of a field with frequency $\omega_i$. With a field strength of $440~(740)$~V/m for microwave field \circled{1}~(\circled{2}) in the main text, we have $(\mathscr{P}_1,\mathscr{P}_2) =-(5.5,1.9)\times10^{-32}m^3$, and an energy shift of about $ 15$~MHz. However, this energy shift happens for all relevant Rydberg states, thus does not alter the vdWI.

\begin{table}
  \begin{tabular}{|c|c|}
    \hline     state  &$\delta E$~(MHz)  \\ \hline
$56p_{3/2}, m=-1/2;57p_{3/2}, m=-1/2 $ &  44  \\
$56p_{3/2}, m=1/2;57p_{3/2}, m=-3/2 $ &   72  \\
$57p_{3/2}, m=1/2;56p_{3/2}, m=-3/2 $&   68  \\
  \hline
  \end{tabular}
  \caption{  \label{table0} The ac Stark shifts for the states involved in the strongest coupled channels from $|\mathsf{n}_1s,\mathsf{n}_1s\rangle$, where $|\mathsf{n}_1s\rangle$ is the state characterized by $57S_{1/2}, m=-1/2$.  }
 \end{table}

\subsection{Off-resonant ac Stark shifts}\label{appCtwo}
Another type of energy shift is due to non-resonant microwave couplings. For instance, $|\mathsf{n}_1s\rangle$ can be coupled off-resonantly to a state $|\mathsf{n}_pp\rangle$ other than $|p\rangle$, causing an ac Stark shift to its energy~\cite{Bohlouli-Zanjani2007}
\begin{eqnarray}
  \delta E_1&=&\sum_{i=1,2}\frac{\mathcal{E}_i^2}{2}\sum_{\mathsf{n}_pp} \frac{( E_{\mathsf{n}_1s}- E_{\mathsf{n}_pp})|\langle \mathsf{n}_1s|\vec{\epsilon}_i\cdot\vec{d}|\mathsf{n}_pp\rangle|^2  }{( E_{\mathsf{n}_1s}- E_{\mathsf{n}_pp})^2- \omega_i^2   },\nonumber\\
  \label{acStark}
\end{eqnarray}
where the sum over a p-orbital state $|\mathsf{n}_pp\rangle$ excludes $|p\rangle$, $\vec{d}$ is the electric dipole operator, $\vec{\epsilon}_i$ the polarization operator of the microwave field, and $\omega_i$ the photon energy of the field with a classical electric field $\mathcal{E}_i$. Numerical calculation with $|\mathsf{n}_p- \mathsf{n}_1|<30$ gives us $\delta E_1=12.4$~MHz and $\delta E_2=4.1$~MHz, for $|\mathsf{n}_1s\rangle$ and $|\mathsf{n}_2s\rangle$, respectively. Comparing to the energy defects $\sim 1.5$~GHz for the channel that contributes strongest, these energy shifts are negligible.

For the Stark shifts of dipole-coupled $p$ levels, we can take the nearest states as a typical example, since they are coupled strongly by dipole interaction and have smallest energy defects, thus contributing significantly to the vdWI. The states that couple strongest to the state $|\mathsf{n}_1s,\mathsf{n}_1s\rangle$ are $ |57p_{3/2},56p_{3/2}\rangle$ and $|56p_{3/2},57p_{3/2}\rangle$, with an energy defect $-1.5$~GHz. The calculation results are in Table~\ref{table0}. Compared to the energy defects $\sim 1.5$~GHz, the ac Stark shifts are negligible. The change of energy defects for other channels is even smaller compared with their energy defects. From these estimates, the energy defects between $|\mathsf{n}_1s,\mathsf{n}_1s\rangle$ and another dipole-coupled state are altered no more than $5\%$. This means that the vdWI of the state $|\mathsf{n}_1s,\mathsf{n}_1s\rangle$ can change in an amount $\lesssim 5\%$. Similar results apply to the vdWI of the states $|\mathsf{n}_2s,\mathsf{n}_2s\rangle$ and $|\mathsf{n}_{2(1)}s,\mathsf{n}_{1(2)}s\rangle$.

\begin{figure}
\includegraphics[width=2.9in]
{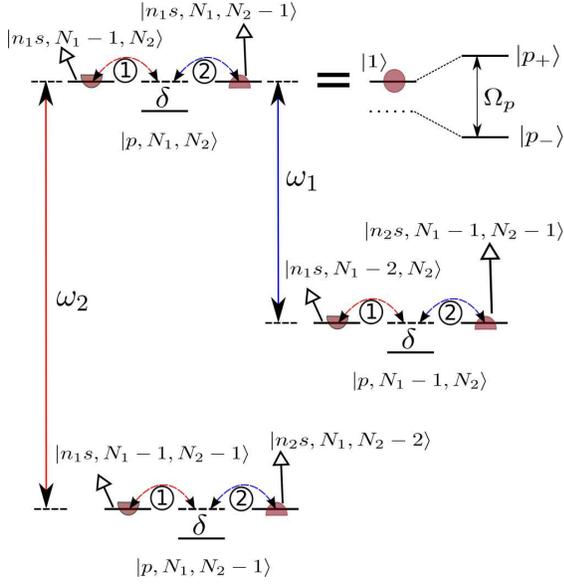}
 \caption{Energy levels of the atom-photon system for a single Rydberg atom with the microwave field dressing.  \label{fig-ac}}
\end{figure}

\subsection{Resonant ac Stark shifts}
Another type of energy shift may arise from the almost resonant couplings, so that the resonant ac Stark effect may alter the vdWI of the state $|11\rangle$. To study this effect, below we describe the microwave fields by quantized operators, so that the effect of microwave field dressing is included while the complexity of introducing time-dependency via a rotating-frame transformation is avoided. In the limit that the applied dressing fields have narrow bands, the classical limit of the Rabi frequencies $\Omega_1$ and $\Omega_2$ are recovered. 

For the example system of the main text, the intermediate state $| p  \rangle$ is located below $|\mathsf{n}_1s\rangle$ and $|\mathsf{n}_2s\rangle$. In this case, the last term of the following Hamiltonian
\begin{eqnarray}
  \hat{H}&=&\hat{H}_{\text{dd}} +\hat{H}_{\text{atom}} + \sum_{i=\text{c,t}}\hat{H}_{\text{M}}, \label{IIC01}
\end{eqnarray}
for each atom is written as
\begin{eqnarray}
  \hat{H}_{\text{M}} &=&\hat{H}_{\text{r}} +  \hat{H}_{\text{nr}}, \nonumber\\
  \hat{H}_{\text{r}}  &=&\sum_{j= n_1s,n_2s,p} E_{j}|j\rangle\langle j|\nonumber\\
  &&+ \sum_{i=1,2}\left[\omega_ia_i^\dag a_i+ \frac{g_i}{2}\left(a_i^\dag | p  \rangle \langle \mathsf{n}_i s | +\text{h.c.}\right) \right], \nonumber
\end{eqnarray}
where $g_i$ is the single-photon coupling strength between the $i$th field and the atom, while $a_i$ and $a_i^{\dagger}$ are annihilation and creation operators, respectively, for the $i$th field. The atomic energies for the three microwave-dressed states have been explicitly written out, and the anti-rotating term that does not conserve energy is
\begin{eqnarray}
\hat{H}_{\text{nr}} &=&\sum_{i=1,2} \frac{g_i}{2}\left(a_i| p  \rangle \langle \mathsf{n}_i s | +\text{h.c.}\right).\nonumber
\end{eqnarray}
We employ a quantized field description of the microwave fields only as a convenient way of presenting a time-independent description; the transformation to a classical field picture is well understood.

We treat the resonant microwave coupling and the atomic energy terms as non-perturbed part of the Hamiltonian $\hat{\mathcal{V}}_{0}$, and other parts as a perturbation $\hat{\mathcal{V}}_{1}$, i.e.,
\begin{eqnarray}
  \hat{H}&=&\hat{\mathcal{V}}_{0} +\hat{\mathcal{V}}_{1}\nonumber\\
  \hat{\mathcal{V}}_{0} &=& \hat{H}_{\text{atom}}+ \hat{H}_{\text{r}} ,\nonumber\\
  \hat{\mathcal{V}}_{1} &=& \hat{H}_{\text{dd}} +\hat{H}_{\text{nr}}.\nonumber
\end{eqnarray}
Now consider the Hamiltonian $ \hat{\mathcal{V}}_{0}$ for the three states $| p,N_1,N_2  \rangle,| \mathsf{n}_1 s,N_1-1,N_2  \rangle,|\mathsf{n}_2 s,N_1,N_2-1  \rangle $, where $N_i$ denotes the number of photons of the $i$th microwave field. With the convention of $E_{n_1s}+ (N_1-1)\omega_1+N_2\omega_2=E_{n_2s}+ N_1\omega_1+ (N_2-1)\omega_2=0$, so that $E_p+ N_1\omega_1+N_2\omega_2=-\delta$, we have
\begin{eqnarray}
  \hat{\mathcal{V}}_{0}
 &= & \hat{H}_{\text{atom}}+ \left( \begin{array}{ccc}
    -\delta & \sqrt{N_1}g_1/2 &  \sqrt{N_2}g_2/2\\
    \sqrt{N_1}g_1/2 & 0 &0 \\
     \sqrt{N_2}g_2/2 &0&0
   \end{array}\right).\nonumber\\ \label{appCV0}
\end{eqnarray}
If the number of photons in the microwave field is centered around $N_i$, $i=1$ or $2$, then
\begin{eqnarray}
  \sqrt{N_1}g_1 &\approx & \Omega\cos\beta,\nonumber\\
  \sqrt{N_2}g_2&\approx & \Omega\sin\beta,\nonumber
\end{eqnarray}
and the microwave field Hamiltonian is then equivalent to that in Eq.~(\ref{HM001}) except of a different sign of the parameter $\delta$. The dark eigen state for the microwave dressing field of Eq.~(\ref{appCV0})
\begin{eqnarray}
  |1\rangle = \sin\beta| \mathsf{n}_1 s,N_1-1,N_2  \rangle - \cos\beta|\mathsf{n}_2 s,N_1,N_2-1  \rangle ,\nonumber\\
  \label{define1withphoton}
\end{eqnarray}
is separated from the other two eigenstates
\begin{eqnarray}
   |p_+\rangle&=&\sin\gamma |+ \rangle +\cos\gamma |p \rangle,\nonumber\\
   |p_-\rangle&=&\cos\gamma |+ \rangle -\sin\gamma |p \rangle,\nonumber
\end{eqnarray}
with energy gaps~[see Eq.~(\ref{eqA10})] that are large compared with the dipole interaction, where
\begin{eqnarray}
   \sin\gamma &=& \Omega/\mathcal{N}_p, \cos\gamma = (\Omega_{\text{p}}+\delta)/\mathcal{N}_p,\nonumber\\
   \mathcal{N}_p&= & \sqrt{\Omega^2+(\Omega_{\text{p}}+\delta)^2},\Omega_{\text{p}}=\sqrt{\Omega^2+\delta^2},\nonumber\\
    |+\rangle&=&\cos\beta | \mathsf{n}_1 s,N_1-1,N_2  \rangle +\sin\beta|\mathsf{n}_2 s,N_1,N_2-1  \rangle .\nonumber
\end{eqnarray}
A picture showing this ac Stark effect is given in Fig.~\ref{fig-ac}.

We are particularly interested in the vdWI of the state $|11\rangle\equiv |1,1\rangle= |1\rangle\otimes |1\rangle$, which can be calculated by the quasi-degenerate perturbation theory~\cite{Shavitt1980},
\begin{eqnarray}
  \hat{H}_{\text{v}} &=& \hat{\Upsilon} \left( \hat{\mathcal{V}}_{1} + \hat{\mathcal{V}}_{1}\hat{G}'\hat{\mathcal{V}}_{1} +\hat{\mathcal{V}}_{1}\hat{G}' \hat{\mathcal{V}}_{1}\hat{G}'\hat{\mathcal{V}}_{1} +\cdots \right)\hat{\Upsilon}\nonumber\\
  &=& \hat{H}_{\text{v}}^{(1)} +\hat{H}_{\text{v}}^{(2)} +  \hat{H}_{\text{v}}^{(3)} +\cdots , \label{GreenFunction}
\end{eqnarray}
where $ \hat{\Upsilon} = |11\rangle\langle 11|$ is the projection operator, $\hat{G}'$ is the Green's function $\hat{G}' = (\hat{1} -  \hat{\Upsilon}) \frac{1}{E_0 -\hat{\mathcal{V}}_{0}  }(\hat{1} -  \hat{\Upsilon})$ which vanishes for the state $|11\rangle$. Obviously $\hat{H}_{\text{v}}^{(1)} =0$. The contribution to $\hat{H}_{\text{v}}^{(2)}$ from $\hat{H}_{\text{dd}}$ not only contains the usual vdWI as the one in the main text~[as can be calculated later on, see Eqs.~(\ref{eqC6})], but also has a contribution from the dipole processes such as
\begin{eqnarray}
 | \mathsf{n}_1 s,  \mathsf{n}_1 s \rangle   \rightarrow  | \mathsf{n}_1' p,  \mathsf{n}_2' p \rangle  \rightarrow  | \mathsf{n}_2 s,  \mathsf{n}_2 s \rangle . \label{eqextra}
\end{eqnarray}
The processes above contribute negligibly to the vdWI of the state $|11\rangle$, as can be shown by the following example, 
\begin{eqnarray}
 | 57 s, 57 s \rangle   \rightarrow  | 57 p_{3/2},  56 p_{3/2} \rangle  \rightarrow  | 59 s, 59 s \rangle.\nonumber
\end{eqnarray}
The two coupling strengths in the two dipole processes above are very different in magnitude: the latter one is more than $100$ times smaller than the former. In other words, the channel above is at least 100 times weaker compared with
\begin{eqnarray}
 | 57 s, 57 s \rangle   \rightarrow  | 57 p_{3/2},  56 p_{3/2} \rangle  \rightarrow  | 57 s, 57 s \rangle.\nonumber
\end{eqnarray}
So we can neglect the extra channels like Eq.~(\ref{eqextra}).

 We write $\hat{H}_{\text{dd}}$ as
\begin{eqnarray}
  \hat{H}_{\text{dd}} &=&\hat{V}_{11} + \hat{V}_{22} +  \hat{V}_{12} + \hat{V}_{21},\nonumber\\
\hat{V}_{11}&=&  V_{11}(|\mathsf{n}_1s, m_1;\mathsf{n}_1s, m_2\rangle \langle\mathsf{n}_{1}'p, m_3;\mathsf{n}_{2}'p, m_4|+\text{h.c.})\nonumber\\
\hat{V}_{22}&=&  (V_{22}|\mathsf{n}_2s, m_1;\mathsf{n}_2s, m_2\rangle \langle\mathsf{n}_{1}'p, m_3;\mathsf{n}_{2}'p, m_4|+\text{h.c.})\nonumber\\
\hat{V}_{12}&=&  (V_{12}|\mathsf{n}_1s, m_1;\mathsf{n}_2s, m_2\rangle \langle\mathsf{n}_{1}'p, m_3;\mathsf{n}_{2}'p, m_4| +\text{h.c.}),\nonumber\\
 && \label{appCV123}
\end{eqnarray}
where the sum over $\mathsf{n}_{1}',\mathsf{n}_{2}'$, and $m_k$, $k=1,\cdots,4$, are not explicitly written out, while $\hat{V}_{21}$ is similar to $\hat{V}_{12}$. The three pieces of dipole operators $\hat{V}_{11}$, $\hat{V}_{22}$ and $\hat{V}_{12},\hat{V}_{21}$ almost operate independently on the state $|11\rangle$ due to the difference of the two principal numbers $\mathsf{n}_1$ and $\mathsf{n}_2$.

The contribution to $\hat{H}_{\text{v}}^{(2)}$ from $\hat{H}_{\text{dd}}$ also arises from the states $|p_\pm p_ \pm\rangle,|p_\pm  p_\mp\rangle$ because of the dipole coupling between $|11\rangle$ and $|pp\rangle$. This dipole coupling strength, however, is small due to the large difference in their principal quantum numbers of the three microwave dressed states. As a consequence, this extra channel does not alter the vdWI contribution to $\hat{H}_{\text{v}}^{(2)}$ from $\hat{H}_{\text{dd}}$. There is also some contribution from $\hat{H}_{\text{nr}}$ to $\hat{H}_{\text{v}}^{(2)}$. As an example,
\begin{eqnarray}
  |11\rangle&& \rightarrow[ |1\rangle \otimes(\sin\beta| p,N_1-2,N_2  \rangle\nonumber\\
  &&- \cos\beta|\mathsf{n}_2 s,N_1,N_2-1  \rangle)+(\sin\beta| p,N_1-2,N_2  \rangle\nonumber\\
  &&- \cos\beta|\mathsf{n}_2 s,N_1,N_2-1  \rangle)\otimes|1\rangle] /\sqrt2,\nonumber
\end{eqnarray}
gives us a perturbation energy,
\begin{eqnarray}
  \frac{\Omega^2\cos^2\beta}{4\omega_1},\nonumber
\end{eqnarray}
i.e., the Bloch-Siegert shift~\cite{CohenTannoudji}, which is smaller than $1$~MHz for the parameters of the main text. So the Bloch-Siegert shift is negligible. Notice that the contribution to $\hat{H}_{\text{v}}^{(2)}$ from the cross terms of $\hat{H}_{\text{dd}}$ and $\hat{H}_{\text{nr}}$ is zero, simply because $\hat{H}_{\text{dd}}$ flips the states of both atoms, while $\hat{H}_{\text{nr}}$ only flips the state of one atom at a time, thus does not give nonzero terms as seen from Eq.~(\ref{GreenFunction}).

Below we calculate the vdWI of the state $|11\rangle=|--\rangle $ given by $\hat{H}_{\text{v}}^{(2)}$. The vdWI from $\hat{V}_{11}$ of Eq.~(\ref{appCV123}) is
\begin{eqnarray}
- \langle --|\hat{V}_{11}| \mathsf{n}_{1}'\mathsf{n}_{2}'\rangle \langle \mathsf{n}_{1}'\mathsf{n}_{2}'| \hat{V}_{11}^\dag |--\rangle / \delta_{\mathsf{n}_{1}'\mathsf{n}_{2}'}, \label{eqC6}
\end{eqnarray}
where 
\begin{eqnarray}
 | \mathsf{n}_{1}'\mathsf{n}_{2}'\rangle &=&(\sin\beta | \mathsf{n}_{1}'p \rangle-\cos\beta |\mathsf{n}_{2}s \rangle)\nonumber\\
 && \otimes(\sin\beta | \mathsf{n}_{2}'p \rangle-\cos\beta |\mathsf{n}_{2}s \rangle),\nonumber
\end{eqnarray}
and the energy defect above is
\begin{eqnarray}
  \delta_{\mathsf{n}_{1}'\mathsf{n}_{2}'} &=& [  (\sin\beta | \mathsf{n}_{1}'p \rangle-\cos\beta |\mathsf{n}_{2}s \rangle)\nonumber\\
    && \otimes(\sin\beta | \mathsf{n}_{2}'p \rangle-\cos\beta |\mathsf{n}_{2}s \rangle)]^{\dag} \otimes\hat{H}_{\text{atom}} \nonumber\\
  &&\otimes (\sin\beta | \mathsf{n}_{1}'p \rangle-\cos\beta |\mathsf{n}_{2}s \rangle)\nonumber\\
 && \otimes(\sin\beta | \mathsf{n}_{2}'p \rangle-\cos\beta |\mathsf{n}_{2}s \rangle)  -E_{--},\nonumber
\end{eqnarray}
where $E_{--}=0$ as defined above Eq.~(\ref{appCV0}). Here we have not explicitly written out the photon states. Obviously, the energy defect $\delta_{\mathsf{n}_{1}'\mathsf{n}_{2}'}$ is simply the energy defect of the following process
\begin{eqnarray}
 |\mathsf{n}_{1}s \mathsf{n}_{1}s \rangle\rightarrow| \mathsf{n}_{1}'p \mathsf{n}_{2}'p \rangle\nonumber
\end{eqnarray}
times a factor of $\sin^4\beta$. In a similar way, there is a factor of $\sin^8\beta$ in the numerator of Eq.~(\ref{eqC6}), as a consequence, the result of Eq.~(\ref{eqC6}) gives us
\begin{eqnarray}
C_6(\mathsf{n}_{1},\mathsf{n}_{1})\sin^4\beta/L^6  = a\sin^4\beta/L^6  ,\nonumber
\end{eqnarray}
where $a$ is defined below Eq.~(\ref{r1definition}).

Following the same procedure above, we can calculate the vdWI from $\hat{V}_{22},\hat{V}_{12}$ of Eq.~(\ref{appCV123}), giving $d\cos^4\beta/L^6$ and
$(b+c)\sin^2\beta\cos^2\beta/L^6$, respectively. Similarly we can calculate the contribution from $\hat{V}_{21}$. This completes the calculation for the vdWI of the state $|--\rangle$.

\subsection{Residual dipole-dipole couplings upon $|11\rangle$}
Because $|--\rangle$ and $|-p_+\rangle, \cdots$ are not degenerate, they can only be coupled from $|--\rangle=|11\rangle$ by a first-order dipole interaction. Therefore,
\begin{eqnarray}
  |-p_+\rangle, |p_+-\rangle,  |-p_-\rangle, |p_--\rangle\nonumber
\end{eqnarray}
will not be coupled by dipole interaction because the parities of both dipole-coupled atoms should change. However, intuitively, the states
\begin{eqnarray}
  |p_+p_+\rangle ,
  |p_-p_-\rangle,
  |p_+p_-\rangle ,|p_-p_+\rangle\nonumber
\end{eqnarray}
seem to be coupled with $|11\rangle$ through the following four processes,
\begin{eqnarray}
  |\mathsf{n}_1s;\mathsf{n}_1s \rangle\rightarrow |p;p \rangle,~~30.0\text{MHz}\mu m^3 \mathcal{M}/L^3,\nonumber\\
  |\mathsf{n}_1s;\mathsf{n}_2s \rangle\rightarrow |p;p  \rangle,~~11.4\text{MHz}\mu m^3 \mathcal{M}/L^3\nonumber\\
  |\mathsf{n}_2s;\mathsf{n}_1s \rangle\rightarrow |p;p  \rangle,~~11.4\text{MHz}\mu m^3 \mathcal{M}/L^3\nonumber\\
  |\mathsf{n}_2s;\mathsf{n}_2s \rangle\rightarrow |p;p  \rangle,~~4.35\text{MHz}\mu m^3 \mathcal{M}/L^3.\label{strongDiple}
\end{eqnarray}
 Nevertheless, each state on the left hand side of Eq.~(\ref{strongDiple}) has a total spin of $-1/2-1/2=-1$, but $|p;p \rangle$ has a total spin of $-3/2-3/2=-3$, thus the processes above could not happen due to the conservation of angular momentum. So, no coupling happens between $|--\rangle$ and other states. 

\begin{figure}
\includegraphics[width=3.3in]
{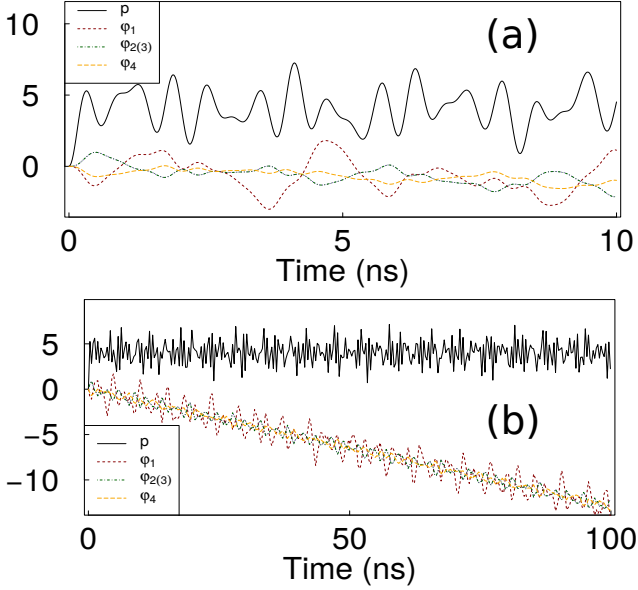}
\caption{Numerical evidence of the dark state $|11\rangle$ with $(\mathsf{n}_1,\mathsf{n}_2)=(57,59)$, and the two-atom distance $L=4.4\mu$m. Here $p = 10^4(1-|\langle \psi(t)|11\rangle|^2)$ is the population error of the dark state $|11\rangle$ scaled by $10^4$, while $\varphi_i/100$, where $i=1,\cdots,4$, represents the phase of the state $|11\rangle$. (a) and (b) present the short and long time scale results, respectively. \label{fig-dark}}
\end{figure}

\section{Numerical simulation as a test for the presence of the dark state $|11\rangle$}\label{app_C}

We perform a numerical simulation to confirm the robustness of the dark state $|11\rangle$ as follows: the dark state $|11\rangle$ is populated in the presence of the dressing microwave field, and we calculate its subsequent time evolution to check that it is indeed a dark state. Furthermore, by calculating a phase parameter we show that the vdWI of the state $|11\rangle$ is zero.  

The state $|11\rangle$ can be expanded by using Eq.~(\ref{define1withphoton}),
\begin{eqnarray}
  |11\rangle &=& \sin^2\beta| \mathsf{n}_1 s,N_1-1,N_2  \rangle| \mathsf{n}_1 s,N_1-1,N_2  \rangle \nonumber\\
  &&- \sin\beta\cos\beta (| \mathsf{n}_1 s,N_1-1,N_2  \rangle|\mathsf{n}_2 s,N_1,N_2-1  \rangle \nonumber\\
  && +|\mathsf{n}_2 s,N_1,N_2-1  \rangle | \mathsf{n}_1 s,N_1-1,N_2  \rangle )\nonumber\\
  && +\cos^2\beta |\mathsf{n}_2 s,N_1,N_2-1  \rangle |\mathsf{n}_2 s,N_1,N_2-1  \rangle ,\nonumber
\end{eqnarray}
where the first~(second) ket denotes the state for the first~(second) qubit. Now the microwave field interaction with each qubit is given by the Hamiltonian of Eq.~(\ref{appCV0}), while a dipolar interaction couples the two-atom state with other states. Because the dipole-dipole interaction changes atomic states, we can suppress the photonic arguments in the equation above, and then  
\begin{eqnarray}
  |11\rangle &=& \sin^2\beta| \mathsf{n}_1 s  \rangle| \mathsf{n}_1 s  \rangle - \sin\beta\cos\beta (| \mathsf{n}_1 s  \rangle|\mathsf{n}_2 s  \rangle \nonumber\\
  && +|\mathsf{n}_2 s  \rangle | \mathsf{n}_1 s  \rangle ) +\cos^2\beta |\mathsf{n}_2 s \rangle |\mathsf{n}_2 s  \rangle . \label{11expand}
\end{eqnarray}
Dipole interaction will generate the following couplings, 
\begin{eqnarray}
  &&| \mathsf{n}_1 s  \rangle| \mathsf{n}_1 s  \rangle\mapsto | \mathsf{n}_1' p  \rangle| \mathsf{n}_1'' p  \rangle, \nonumber\\
  &&| \mathsf{n}_1 s  \rangle|\mathsf{n}_2 s  \rangle\mapsto | \mathsf{n}_1' p  \rangle| \mathsf{n}_2'' p  \rangle, \nonumber\\
  && |\mathsf{n}_2 s  \rangle | \mathsf{n}_1 s  \rangle \mapsto | \mathsf{n}_2' p  \rangle| \mathsf{n}_1'' p  \rangle, \nonumber\\
  &&|\mathsf{n}_2 s \rangle |\mathsf{n}_2 s  \rangle\mapsto | \mathsf{n}_2' p  \rangle| \mathsf{n}_2'' p  \rangle .\label{IIIeq03}
\end{eqnarray}
Note that if the atomic parts of two states on the right hand side of two different lines above are the same, they still represent different states. This is because the photon states on each line above are different from that of another line. We will include the $5^2\times8$ channels that satisfy the condition of $|\mathsf{n}_1-\mathsf{n}_1'(\mathsf{n}_1'')|, |\mathsf{n}_2-\mathsf{n}_2'(\mathsf{n}_2'')|\leq 2$, since other channels have much larger energy defects, and contribute much less to the dynamics~\cite{Walker2008}. Here $5^2$ accounts for the number of combinations of principal quantum numbers $\mathsf{n}_1'  \mathsf{n}_1'' ,\cdots$, while each set of states $| \mathsf{n}_1' p  \rangle| \mathsf{n}_1'' p  \rangle$ contains eight fine structure states: $| \mathsf{n}_1' p_{1/2},-1/2 \rangle| \mathsf{n}_1'' p_{1/2},-1/2  \rangle$, $ | \mathsf{n}_1' p_{3/2},-3/2 \rangle| \mathsf{n}_1'' p_{1/2},1/2  \rangle$, $\cdots$, $| \mathsf{n}_1' p_{3/2},1/2 \rangle| \mathsf{n}_1'' p_{3/2},-3/2  \rangle $. Because dipole interaction conserves the total angular momentum, but all the states on the left hand side of Eq.~(\ref{IIIeq03}) have total electron spin of $-1$, while $|p\rangle|p\rangle$ has total electron spin of $-3$, we do not need to include the extra channels in Eq.~(\ref{strongDiple}). 

The dark state is established by the following couplings with Rabi frequencies that are large compared to any dipole-dipole interaction,
\begin{eqnarray}
  &&| \mathsf{n}_1 s  \rangle| \mathsf{n}_1 s  \rangle\mapsto | \mathsf{n}_1s  \rangle| p  \rangle, | p  \rangle| \mathsf{n}_1s  \rangle, \nonumber\\
  &&| \mathsf{n}_1 s  \rangle|\mathsf{n}_2 s  \rangle\mapsto | \mathsf{n}_1s  \rangle| p  \rangle, | p  \rangle| \mathsf{n}_2s  \rangle , \nonumber\\
  && |\mathsf{n}_2 s  \rangle | \mathsf{n}_1 s  \rangle \mapsto | \mathsf{n}_2s  \rangle| p  \rangle, | p  \rangle| \mathsf{n}_1s  \rangle, \nonumber\\
  &&|\mathsf{n}_2 s \rangle |\mathsf{n}_2 s  \rangle\mapsto  | \mathsf{n}_2s  \rangle| p  \rangle, | p  \rangle| \mathsf{n}_2s  \rangle .
\end{eqnarray}
In principle, we shall include higher-order processes such as $| \mathsf{n}_1s,N_1-1,N_2  \rangle| p ,N_1  ,N_2  \rangle  \mapsto  | \mathsf{n}_1'p,N_1-1,N_2  \rangle| \mathsf{n}_xs(d) ,N_1  ,N_2  \rangle  $. Nevertheless, since the population of the states on the right hand side of the equations above is negligible, we will neglect these higher-order terms in a first approximation. Including the original four basis states of $|11\rangle$ in Eq.~(\ref{11expand}), all the states coupled by dipole interaction and those coupled by the dressing microwave fields, there are $208$ states in our numerical calculation to verify the dark state $|11\rangle$.

We proceed as follows. Populate the initial two-atom state $|11\rangle$, then let the wave function $|\psi(t)\rangle$ evolve under the control of the Hamiltonian in Eq.~(\ref{IIC01}), and then compute the following five values during the time evolution, 
\begin{eqnarray}
  p&=& 10^4(1-|\langle \psi(t)|11\rangle|^2), \nonumber\\
    \varphi_1&=&10^2 \text{angle}[ \langle\psi(t) | \mathsf{n}_1 s  \rangle| \mathsf{n}_1 s  \rangle],\nonumber\\
    \varphi_2 &=&10^2 \text{angle}[ -\langle\psi(t) | \mathsf{n}_1 s  \rangle| \mathsf{n}_2 s  \rangle],\nonumber\\
    \varphi_3 &=&10^2 \text{angle}[- \langle\psi(t) | \mathsf{n}_2 s  \rangle| \mathsf{n}_1 s  \rangle],\nonumber\\
    \varphi_4 &=&10^2 \text{angle}[ \langle\psi(t) | \mathsf{n}_2 s  \rangle| \mathsf{n}_2 s  \rangle],
\end{eqnarray} 
where angle$(f)$ gives the argument of a complex number. In the ideal limit the microwave Rabi frequencies $\Omega_{1(2)}$ are infinitely large compared to any dipole coupling, so that $10^{-4}p$ remains zero. Its small fluctuation around zero is due to the finite $\Omega_{1(2)}$ in this work. The numerical results for both short and long time scales are shown in Fig.~\ref{fig-dark}. In both Fig.~\ref{fig-dark}(a) and (b) the population error remains bounded. The error is centered around $4\times10^{-4}$ and provides direct evidence of the robustness of the dark state $|11\rangle$.

Besides the population conservation there is further evidence of the dark state's robustness: in case of nonzero $\overline{C}_6$ for the dark state, each of the four phases $\varphi_i$, where $i=1,\cdots,4$, should grow linearly in time according to $Vt$, where $V=\overline{C}_6/L^6$. This is shown in Fig.~\ref{fig-dark}(b) in the long time scale dynamics. Here we find that there is an overall phase accumulation $Vt$ of the state $|11\rangle$, where $V\approx0.22$~MHz by a linear fitting. The appearance of such residual vdWI is because, as shown following Eq.~(\ref{IIIeq03}), we only include the dipole-dipole interaction channels that satisfy the condition of $|\mathsf{n}_1-\mathsf{n}_1'(\mathsf{n}_1'')|, |\mathsf{n}_2-\mathsf{n}_2'(\mathsf{n}_2'')|\leq 2$ in the numerical simulation. As a result, the C-six coefficients deviate a little from those in the main text~(where many more channels are included), leading to an effective C-six coefficient $\overline{C}_6\approx2107$~MHz$\mu m^6$. Such a residual C-six coefficient gives us $\overline{V}=0.29$~MHz, quite near to the value $V\approx0.22$~MHz from Fig.~\ref{fig-dark}(b). The discrepancy can be understood by the fact that, for the sake of simplicity, we have ignored the processes like $|\mathsf{n}_2 s  \rangle| \mathsf{n}_1 s  \rangle\mapsto | \mathsf{n}_1' p  \rangle| \mathsf{n}_1'' p  \rangle\mapsto| \mathsf{n}_1 s  \rangle|\mathsf{n}_2 s  \rangle$ in Eq.~(\ref{IIIeq03}), i.e., the dipole interaction that exchanges the two principal quantum numbers~[which gives rise to $b$ following Eq.~(3) of the main text]. In principle, inclusion of such channels should give a result nearer to the theoretical value of $\overline{V}=0.29$~MHz.

\begin{figure}
\includegraphics[width=2.5in]
{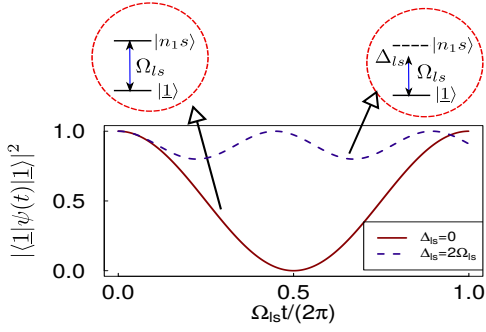}
\caption{Population of $|\underline{1}\rangle$ as a function of time $t$ when a single atom is excited to the Rydberg state with Rabi frequency $\Omega_{\text{ls}}$ and detuning $\Delta_{\text{ls}}$. The initial state is $|\underline{1}\rangle$.\label{fig-f1}}
\end{figure}

\begin{figure}
\includegraphics[width=2.5in]
{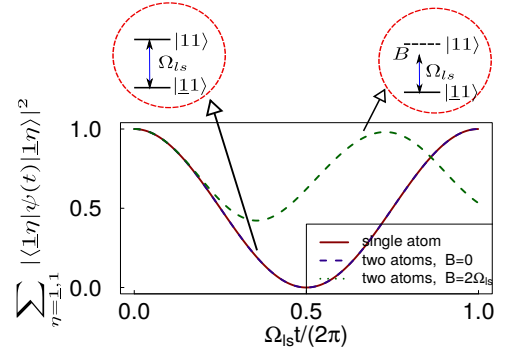}
\caption{Sum of populations on the states $|\underline{1}\underline{1}\rangle$ and $|\underline{1}1\rangle$ as a function of time $t$ when each of the two interacting atoms are excited to the Rydberg state $|1\rangle$ with a Rabi frequency $\Omega_{\text{ls}}$ and two-atom Rydberg blockade $B$. For $B=0$, the ground state population of one atom has the same time evolution as that of an isolated atom, while for $B\neq 0$, the time evolution deviates from the single-atom case. The initial state is $|\underline{1}\rangle(|\underline{1}\underline{1}\rangle)$ for a one~(two) atom system.\label{fig-f2}}
\end{figure}

\section{Procedure to locate the mixing angle $\beta_0$ for zero vdWI}\label{app_D}
Because of some extra channels like those in Eq.~(\ref{eqextra}), and especially because there are ac Stark shifts for the Rydberg levels, the vdWI coefficients $a,b,c$, and $d$ may change a little bit, for instance, up to $5\%$ as discussed in Sec.~\ref{appCtwo}. In this case, the predicted value of $\beta_0$ without accounting for the ac Stark shift of microwave field may result of nonzero vdWI. Thus it is necessary to tune the microwave field strengths so that zero vdWI arises as follows. 

\noindent 1. Turn on the microwave dressing.

\noindent 2. Tune the lasers to be resonant with the transition between the Rydberg states and the ground state $|\underline{1}\rangle$. Notice that the initial laser frequency should be taken near to the transition frequency between the Rydberg state and the ground state $|\underline{1}\rangle$. Take $ |\mathsf{n}_1s\rangle$ as an example, we do not excite $|\mathsf{n}_2s\rangle$, but only $|\mathsf{n}_1s\rangle$. The resonant condition $\Delta_{\text{ls}}=0$ can be tested by measuring the population of the ground state $|\underline{1}\rangle$: if it can not reach zero, then $\Delta_{\text{ls}}\neq0$, but if it can reach zero, then $\Delta_{\text{ls}}=0$, as shown in Fig.~\ref{fig-f1} with two values of $\Delta_{\text{ls}}$ as an example. When resonance is established, the Hamiltonian for one atom is
\begin{eqnarray}
  \hat{H}_{\text{test}}&=& \frac{\Omega_{\text{ls}}' }{2 }\big[| \underline{1}\rangle \left(  \sin\beta  \langle 1| + \cos\beta \sin\gamma \langle p_+| \right. \nonumber\\
  &&+\cos\beta \cos\gamma \langle p_-| \big) +\text{h.c.}\big] + \hat{H}_{\text{M}},\nonumber
\end{eqnarray}
where $\gamma$ and $\hat{H}_{\text{M}}$ are defined in Eqs.~(\ref{pppm}) and ~(\ref{eqA4}), respectively. When $\Omega_{\text{ls}}'\ll |\pm \Omega_{\text{p}}  +\delta|/2$, the coupling between $|\underline{1}\rangle$ and $|p_\pm\rangle$ is suppressed. Define $\Omega_{\text{ls}} =\sin\beta \Omega_{\text{ls}}'$, then
\begin{eqnarray}
  \hat{H}_{\text{test}}& \approx & \frac{\Omega_{\text{ls}} }{2 }\big[| \underline{1}\rangle   \langle 1| +\text{h.c.}\big] + \hat{H}_{\text{M}}.\nonumber
\end{eqnarray}

\noindent 3. The third step is to test if the current microwave field configuration gives a dark state $|11\rangle$ with zero vdWI. Let two atoms trapped close to each other so that vdWI arises. By using the resonant laser pumping between the ground state $|\underline{1}\rangle$ and $|\mathsf{n}_1s\rangle$, and when there is blockade $B$ between the two atoms, the excitation from $|1 \underline{1}\rangle$ and $|\underline{1}1\rangle$ to $|11\rangle$ is shifted away from resonance, thus the ground state population of either atom can not reach zero; take the population of the ground state for the first atom as an example, this is shown by the dashed curve in Fig.~\ref{fig-f2}; When $B=0$, the ground state population of either atom can reach zero, and has exactly the same pattern as an isolated atom, as shown by the solid curve in Fig.~\ref{fig-f2}. By using a small enough $\Omega_{\text{ls}}$, the test result that either atom can reach zero ground state population during one Rabi cycle tells us that we have $B\ll \Omega_{\text{ls}}$. This may give us a $B$ as small as possible.

\noindent 4. If the test of the third step fails, one goes to the first step, tune the microwave field strength so that $\beta$ is changed toward $\beta_0$.

\section{Procedure to set the two lasers in the Y configuration}\label{app_E}
After the correct mixing angle $\beta_0$ is found, we need to find the correct strength ratio of the two lasers in the two legs of the Y configuration, so that only the state $|1\rangle$ is coupled, see Fig.~\ref{fig001} of the main text. The correct laser configuration results of,
\begin{eqnarray}
(\overline\Omega_1,\overline\Omega_2) = (\sin\beta_0,-\cos\beta_0)\overline\Omega ,
\end{eqnarray}
where $\overline{\cdots}$ denotes a two-photon Rabi frequency from laser-atom couplings. Notice that the phase of one laser relative to the other laser also matters in this case: their relative phase shall have a $\pi$ difference to that of the two microwave fields. Below we describe methods about how to set the relative phase and strengths of the two lasers. We assume that we have switched on the microwave dressing with the correct configuration for the mixing angle $\beta_0$.

\subsection{Procedure to set the correct relative phase of the two lasers in the Y configuration}
The correct phase difference between the two lasers in the two legs of the Y configuration is $\pi$. But let us assume that there is a little deviation $\phi_{\text{ls}}$ from this phase, i.e.,
\begin{eqnarray}
\langle  \mathsf{n}_1s|\hat{H}_{\text{ls}} |\underline{1}\rangle &=& \kappa_{n_1}'/2, \nonumber\\
\langle  \mathsf{n}_2s|\hat{H}_{\text{ls}} |\underline{1}\rangle &=& -\kappa_{n_2}'e^{i\phi_{\text{ls}}}/2,\nonumber
\end{eqnarray}
where $\kappa_{n_1}$ and $\kappa_{n_2}$ are real and have the same sign. Now independently switch on $\kappa_{n_1}'$ and $-\kappa_{n_2}'e^{i\phi_{\text{ls}}}$, observe the times for one Rabi cycle of the ground state, which are
\begin{eqnarray}
\tau_{n_1} &=&2\pi/ |\kappa_{n_1}'\sin\beta_0|, \nonumber\\
\tau_{n_2} &=& 2\pi/|\kappa_{n_2}'\cos\beta_0|.\nonumber
\end{eqnarray}
After this, we may simultaneously turn on $\kappa_{n_1}'$ and $-\kappa_{n_2}'e^{i\phi_{\text{ls}}}$, then the time for one Rabi cycle of the ground state becomes
\begin{eqnarray}
\tau_{n_1,n_2} &=&2\pi/\kappa' , \nonumber\\
\kappa' &=&[ (-\kappa_{n_2}'\cos\beta_0\cos\phi_{\text{ls}} +\kappa_{n_1}'\sin\beta_0)^2 \nonumber\\
  &&+(\kappa_{n_2}'\cos\beta_0\sin\phi_{\text{ls}})^2]^{1/2}.\nonumber
\end{eqnarray}
Obviously,
\begin{eqnarray}
1/\tau_{n_1,n_2} &\leq &1/ \tau_{n_1}+1/ \tau_{n_2},\nonumber
\end{eqnarray}
and the equality is true only when $\phi_{\text{ls}}=0$. This can guide us to tune the laser phases to the correct configuration.

\subsection{Procedure to locate the correct strength ratio of the two lasers in the Y configuration}
\noindent 1. Locate the value of $|\Omega_{\text{p}}  +\delta|/2$ for tuning the laser frequency: from being resonant with  $|\mathsf{n}_1s\rangle$ and $|\mathsf{n}_2s\rangle$ to resonance with the state $|p_+\rangle$ for one atom.

\noindent 2. Analytically estimate the laser Rabi frequencies according to the laser output, adjust the ratio of the powers of the two lasers on the two legs in the Y configuration, so that their power ratio is near to the wanted value.

\noindent 3. Switch on both lasers upon the two closely placed atoms, tune the frequencies of both lasers in the two legs of the Y configuration upward simultaneously by a value of $|\Omega_{\text{p}}  +\delta|/2$, while maintaining their powers. By tuning the frequencies of both lasers simultaneously little by little up and down, if no resonance happens between the ground state and the Rydberg state, the strengths of the two lasers have the correct configuration. If there is resonance between the ground state and the Rydberg state, then the laser configuration is incorrect. The reason is as follows. Suppose $\hat{H}_{\text{ls}}$ is the Hamiltonian for the coupling between lasers and atoms,
\begin{eqnarray}
\langle  \mathsf{n}_1s|\hat{H}_{\text{ls}} |\underline{1}\rangle &=& \kappa_{n_1}/2, \nonumber\\
\langle  \mathsf{n}_2s|\hat{H}_{\text{ls}} |\underline{1}\rangle &=& \kappa_{n_2}/2,\nonumber
\end{eqnarray}
then from Eqs.~(\ref{pppm}) and (\ref{jiajian1}),
\begin{eqnarray}
\langle  p_+ |\hat{H}_{\text{ls}} |\underline{1}\rangle &=&\sin\gamma (\cos\beta_0 \kappa_{n_1} +\sin\beta_0 \kappa_{n_2})/2 , \nonumber\\
\langle  p_-|\hat{H}_{\text{ls}} |\underline{1}\rangle &=& \cos\gamma (\cos\beta_0 \kappa_{n_1} +\sin\beta_0 \kappa_{n_2})/2 .\nonumber
\end{eqnarray}
When the strengths of the two lasers satisfy
\begin{eqnarray}
\cos\beta_0 \kappa_{n_1} +\sin\beta_0 \kappa_{n_2}=0,\nonumber
\end{eqnarray}
 there will be no resonant coupling between $|\underline{1}\rangle$ and $| p_\pm\rangle$ even if the laser frequencies are resonant with the transition frequencies. But if the condition above is not satisfied, we shall tune $\kappa_{n_1} / \kappa_{n_2}$ toward the condition $ (\kappa_{n_1} , \kappa_{n_2}) =(\sin\beta_0,-\cos\beta_0)\overline\Omega $.

\section{Atomic state evolution during the gate sequence} \label{App_F}
We set the time at the beginning of pulse 1 as the starting time. The first pulse is a $\pi$ pulse for the microwave field \circled{i}, with the Hamiltonian
\begin{eqnarray}
\hat{H}_{1} &=&\Omega_1\sum_{i=\text{c,t}}[(|\mathsf{n}_1s\rangle\langle p|+\text{h.c.})/2]_i.\label{030201}
  \end{eqnarray}
The single-atom wavefunction $\alpha_1|\mathsf{n}_1s\rangle + \alpha_2|\mathsf{n}_2s\rangle + \alpha_3|p\rangle$ has an initial state of
\begin{eqnarray}
(\alpha_1, \alpha_2,\alpha_3)& = &\left(\sin\beta  ,-\cos\beta,0\right),\label{030202}
\end{eqnarray}
and its time evolution is given by solving Eqs.~(\ref{030201}) and (\ref{030202})~[i.e., $i\partial_t(\alpha_1, \alpha_2,\alpha_3)=\cdots $],
\begin{eqnarray}
(\alpha_1, \alpha_2,\alpha_3)& = &\sin\beta\left( \cos(\Omega_1t/2), -\cot\beta, -i\sin(\Omega_1t/2) \right),\nonumber
  \end{eqnarray}
which turns out to be
\begin{eqnarray}
(\alpha_1, \alpha_2,\alpha_3)& = &\left( 0, -\cos\beta, -i\sin\beta \right),\label{030205}
  \end{eqnarray}
at the end of the first pulse.

Second, we apply a $2\beta$ pulse for the microwave field \circled{ii} whose Hamiltonian is
\begin{eqnarray}
\hat{H}_{2} &=&\Omega_2\sum_{i=\text{c,t}}[(|\mathsf{n}_2s\rangle\langle p|+\text{h.c.})/2]_i.\label{030204}
  \end{eqnarray}
Solving Eqs.~(\ref{030205}) and (\ref{030204}), we have
\begin{eqnarray}
(\alpha_1, \alpha_2,\alpha_3)& = & \left(0, -\cos f, i\sin f \right),\nonumber
  \end{eqnarray}
where $f=-\beta +\Omega_2(t-\pi/\Omega_1)/2$, which turns out to be
\begin{eqnarray}
(\alpha_1, \alpha_2,\alpha_3)& = &( 0, -1, 0),\label{030206}
  \end{eqnarray}
at the end of the second pulse. Note that the duration of the second pulse is $2\beta/\Omega_2$.

Third, after the waiting period $T$, a third pulse for microwave \circled{ii} is applied, and the Hamiltonian is identical to Eq.~(\ref{030204}). The time evolution is
\begin{eqnarray}
(\alpha_1, \alpha_2,\alpha_3)& = & \left(0, -\cos f_2, i\sin f_2 \right),\nonumber
  \end{eqnarray}
where $f_2=\Omega_2(t-\pi/\Omega_1-2\beta/\Omega_2-T)/2$, which turns out to be
\begin{eqnarray}
(\alpha_1, \alpha_2,\alpha_3)& = &( 0, -\cos\beta, i\sin\beta),\label{030208}
  \end{eqnarray}
at the end of the third pulse. Note that the duration of this pulse is also $2\beta/\Omega_2$.

Finally, a $\pi$ pulse for microwave \circled{i} is applied, with the Hamiltonian of (\ref{030201}), solving Eqs.~(\ref{030201}) and (\ref{030208})
\begin{eqnarray}
(\alpha_1, \alpha_2,\alpha_3)& = &\sin\beta\left(\sin(\Omega_1t'/2) , -\cot\beta, i\cos(\Omega_1t'/2) \right),\nonumber
\end{eqnarray}
where $t'= t-\pi/\Omega_1-4\beta/\Omega_2-T$. The final state will be
\begin{eqnarray}
(\alpha_1, \alpha_2,\alpha_3)& = &\left( \sin\beta, -\cos\beta,0 \right),\nonumber
  \end{eqnarray}
at the end of the fourth pulse. The duration of the final pulse is $\pi/\Omega_1$. The total gate time is $2\pi/\Omega_1+4\beta/\Omega_2+T$. From the analysis above, we see that the time evolution of the last two pulses is exactly the time-reversal process of the first and second pulses.

\section{Method of error estimation}\label{App_G}
The method of error estimation follows from Ref.~\cite{Zhang2012}~(and references therein). To stabilize the superposition state $|--\rangle$, we choose $|\delta|=0.26$~GHz before the gate sequence; this choice of $\delta$ is from an optimization of the gate fidelity when $L=4.4\mu$ m. As discussed later on, ionization~\cite{CohenTannoudji} does not happen with this choice. The fields are right-hand polarized~\cite{Chang1994}, so that the $p_{1/2}$ manifold near $|p\rangle$ does not come in due to angular momentum conservation when $|1\rangle$ has $m_J=-1/2$, as explained in the caption of Fig.~\ref{fig001}. 

The first class of error happens for $\{|01\rangle, |10\rangle, |11\rangle\}$, including spontaneous decay of the atomic states and population leakage due to unwanted two-photon or one-photon transitions. The error from the decay of the Rydberg states is estimated as $E_{\text{de}} \approx T_{\text{g}}/\tau$, where the Rydberg lifetime $\tau$ can be taken as $89.3\mu$s from the lifetime of the state $57S_{1/2}$ at temperature $T=300$~K~\cite{Beterov2009}, while the gate time $T_{\text{g}} = \pi/\mathbb{B} +2(\pi/\Omega_1+2\beta/\Omega_2)$ is approximately $37$~ns. Two types of population leakage happens. First, microwave fields can couple the states $|1\rangle$, $|p\rangle$, and $|2\rangle$ with some nearby levels shown in Fig.~\ref{fig003} via twelve one-photon channels, $k=1,\cdots,12$, marked by the thin curves~(lines) with their respective transition frequencies given in Fig.~\ref{fig003}, where it is noted that the two states $p_{1(3)/2}$ are almost degenerate. This gives an error $E_{\text{one}} \approx\sum\eta_{k} \underline{\Omega}_k^2/(2\Delta_k^2)$, where $\eta_{k}$ and $\underline{\Omega}_j$ are the relevant population and the Rabi frequency for the $j$th channel, respectively. Here the leakage happens via both the stabilization fields and the fields during the gate sequence, which are calculated separately. Second, there is a two-photon resonant process, denoted by two thin dashed lines in Fig.~\ref{fig003}, between $|\mathsf{n}_1s\rangle$ and $|\mathsf{n}_2s\rangle$ via two states $62p_{1(3)/2}$ with a detuning $|\Delta_0|\approx7$~GHz, thus $|1\rangle$ and $|p_\pm\rangle$ are coupled with a Rabi frequency $\Omega^{(2)}$, contributing $E_{\text{two}} \approx2[\Omega^{(2)}\sin\gamma/(\delta+\overline\Omega)]^2 + 2[\Omega^{(2)}\cos\gamma/(\delta-\overline\Omega)]^2$ to the gate error. Errors from other two-photon processes are negligible due to large detunings.

The second type of error occurs only for $|11\rangle$ because of the blockade effect of vdWI and atomic motion. The transition from $|11\rangle$ to $|22\rangle$ could be imperfect due to the blockade shift of $|22\rangle$, which happens during the two $2\beta$ pulses and could be estimated as $E_{\text{bl}}\approx2\mathbb{B}^2\Omega_2^{-2}\sin^2\beta$, where the sin factor characterizes the transferred population that matters with blockade. The vdWI and dipole exchange between $s$ and $p$ states, and the vdWI of two $p$-orbital atoms have much smaller rates, thus can be neglected. Another error for the basis state $|11\rangle$ might be from  a mechanical force~\cite{PhysRevLett.85.2208} between the two atoms in the state $|22\rangle$, which is $F(r)= 6\hbar C_6/r^7$. However, beginning with two atoms with no relative motion, one gate cycle will only add a relative speed $\sim 10^{-3}\mu m/\mu s$, and a separation $\sim10^{-5}\mu m\ll L$. Thus this motion effect can be neglected.

The total error for the gate fidelity is thus given by
\begin{eqnarray}
E_0 &=& E_{\text{de}}+E_{\text{one}}+E_{\text{two}} +E_{\text{bl}}.\nonumber
\end{eqnarray}
With the aforementioned parameters, we have $\{E_{\text{de}},E_{\text{one}},E_{\text{two}} ,E_{\text{bl}} \}\approx\{0.41,2.80,0.26,1.15 \}\times10^{-3}$, leading to a total gate error of about $4.6\times10^{-3}$. 

The results shown in Fig.~\ref{fig004} are from each numerically found minimal error $E_0$ when an appropriate $\Omega$ is used for each given $\mathbb{B}$. Here we have fixed the value of $\delta$ so that any multi-photon resonance does not come in. When varying $\Omega$, the eigenvalues in Eq.~(\ref{eqA10}) change. To make sure that the states in Eq.~(\ref{eqA10}) are separated from $|--\rangle$, we make the following constraint: the smallest eigen-energy in Eq.~(\ref{eqA10}) should be larger than three times of the biggest dipole-dipole interaction between $|\mathsf{n}_{1}s  \mathsf{n}_{1}s \rangle$ and a nearby $|p'p''\rangle$ state, which is smaller than $50$~MHz when $L=4.4\mu m$. This makes sure that the vdWI coupling between the states in Eq.~(\ref{eqA10}) and $|--\rangle$ is suppressed.

\section{Interaction for the intermediate states}\label{App_H}
Since the state $|p\rangle$ is also populated during the gate sequence, there should be some interaction involving it. First of all, there is a resonant dipole interaction as
\begin{eqnarray}
|\mathsf{n}_1s;p\rangle \leftrightarrow|p;\mathsf{n}_1s\rangle,\nonumber\\
|\mathsf{n}_2s;p\rangle \leftrightarrow|p;\mathsf{n}_2s\rangle,\nonumber
\end{eqnarray}
and with the parameters from the main text, the first process above is stronger, which has the following rate
\begin{eqnarray}
0.333\cdot 27.7\text{ MHz }\mu m^3/L^3, \label{eqsp2ps}
\end{eqnarray}
where the factor $0.333$ is due to the angular momentum selection rules. With $L=4.4\mu m$, the rate above is only $0.1$~MHz. While for the interaction between two atoms in the state $|p;p\rangle$, the C-six coefficient is only about $1$~GHz$\mu m^6$, much smaller than that between two atoms of $s$-orbital states with similar principal quantum numbers. The vdWI for the states like $|\mathsf{n}_1s; p\rangle$ are small and negligible, too. As a result, we can neglect the interaction of the intermediate states.

\section{Change of the mixing angle $\beta$ by radiative decay of Rydberg states}\label{App_I}
The mixing angle $\beta$ in Eq.~(\ref{darkstate}) can possibly be changed by different decay rates of the two Rydberg states $|\mathsf{n}_1s\rangle$ and $|\mathsf{n}_2s\rangle$. This can be analyzed as follows.

During the initialization of the qubit states, the decay probability of the $|\mathsf{n}_{1(2)}s\rangle$ state is
\begin{eqnarray}
\vartheta_{1(2)}=\pi/(2\tau_{1(2)}\kappa),
\end{eqnarray}
where $\kappa$ is the laser Rabi frequency for the excitation of the Rydberg state, and $\tau_{1(2)}$ is the lifetime of the $|\mathsf{n}_{1(2)}s\rangle$ state, which is 89.3~(97.2)~$\mu$s at the temperature $T=300$~K. By taking $\kappa/(2\pi)=20$~MHz, we find that there is a probability $\vartheta_ 1-\vartheta_ 2=1.1\times10^{-5}$ for $|\mathsf{n}_{1}s\rangle$ to decay more than $|\mathsf{n}_{2}s\rangle$. This imbalance will give small error to the mixing angle and lead to an extra gate error in the order of $10^{-5}$ that has been ignored in the main text.

During the gate sequence, the microwave pulse is fast, and the main gate time is spent on the wait period, where the state $|\mathsf{n}_{2}s\mathsf{n}_{2}s\rangle$, if the input is $|11\rangle$, can acquire a phase shift. Thus no multiple types of Rydberg excitation exist during this wait period.

From the discussion above, we conclude that the different decay rates of different Rydberg states upon the influence of the mixing angle will be about $1\times10^{-5}$, orders of magnitude smaller than the gate fidelity error, thus can be ignored.

\section{Analysis of ionization}\label{App_J}
Two types of photoionization can happen for a Rydberg atom in the presence of a microwave field, i.e., one-photon or multiphoton ionization~\cite{CohenTannoudji}. The example in the main text may involve the second photoionization above. The ionization energies for states $|\mathsf{n}_{1(2)}s\rangle$ are $  E_i(\mathsf{n}_1) =1133.71$~GHz and $E_i(\mathsf{n}_2) = 1053.99$~GHz, respectively, and the two frequencies of the the microwave fields \circled{1} and \circled{2} are $113.5\text{GHz}-\delta$ and $193.3\text{GHz}-\delta$, respectively. Therefore, the ionization of the state needs at least $1053.99/193.3\sim 6$ photons. To analyze the full ionization dynamics is challenging since there are two different microwave fields. But we can show that the example in the main text does not give ionization. This could be done by studying one transition channel that contributes strongest to ionization. To select this transition starting from $|\mathsf{n}_1s\rangle$, we first look for the nearest transition up from $|\mathsf{n}_1s\rangle$, and find that it should be from field \circled{2} via the level $62p_{1(3)/2}$ and detuning $\delta_1^{(i)}=-7$~GHz. Going from this level upward by one step, we will search for a $d_{3(5)/2}$ level with a detuning $\delta_2^{(i)}$ so that $|\delta_1^{(i)}+\delta_2^{(i)}|$ is smallest. The reason that we look for a $d$-orbital state is because the fields are right-hand polarized, so that the electron spin quantum number increases by one when going up each step. Repeating this process, we can select the channel that contributes strongest to the ionization. Note that, however, quantum interference might decrease its effect when other channels come in~\cite{CohenTannoudji}.
\begin{table}
  \begin{tabular}{|c|c|c|c|c|c|c|}
    \hline     k  & microwave  &  state  &  energy &  transition energy& $\delta_k^{(i)}$ & $\sum_{k'=1}^k\delta_{k'}^{(i)}$  \\ \hline

1&2(1)&    62p&    -933.7&    200.0(120.3)&     -7.0&    -7.0 \\
2&2&    68d&    -740.5&    193.2&     -0.2&    -7.2 \\
3&2&    77f&    -555.1&    185.4&     7.6&    0.4 \\
4&2&    95g&    -364.5&    190.6&     2.4&    2.8 \\
5&1&    115h&    -248.8&    115.8&     -2.5&    0.4 \\
6&1&    156i&    -135.2&    113.6&     -0.3	& 0.1\\
  \hline
  \end{tabular}
  \caption{  \label{table1} The strongest multiphoton transition channel from $|\mathsf{n}_1s\rangle$~($|\mathsf{n}_2s\rangle$). $k$ labels the order of the transition. Here microwave $i$, where $i=1(2)$, means the $i$th field in the main text for stabilizing the Rydberg state $|1\rangle$. The energy of a state refers to one of the two fine states with a larger $J$. For instance, $-933.7$~GHz is the energy of the state $62p_{3/2}$, which is listed for convenience since the fine structure splitting here is small. All the energy is in unit of GHz.  }
 \end{table}

The quantum defects used to calculate the energy of a Rydberg atom are taken from Refs.~\cite{Han2009,Li2003} for $s,p,d$ and $f$-orbital states. Since data for higher orbital states are not available, we ignore the effects of quantum defect because the quantum defects for higher angular momentum states are small. For the example of the main text, the search result with the criteria $|\sum_{k'=0}^k\delta_{k'}^{(i)}|$ being smallest for each step is listed in Tables~\ref{table1} for the transition from $|\mathsf{n}_1s\rangle$ and $|\mathsf{n}_2s\rangle$, where only the first step differs for these two cases.

\begin{figure}
\includegraphics[width=3.1in]
{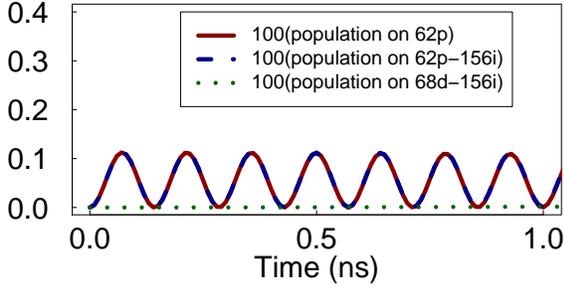}
\caption{Simulated population, scaled by $100$, on the level $62p$, the levels $\{62p,68d,77f,\cdots,156i\}$, and the levels $\{68d,77f,\cdots,156i\}$ as a function of time starting from the initial state $\sin\beta|\mathsf{n}_1s\rangle$. \label{fig005}}
\end{figure}

From Table~\ref{table1}, it is obvious that if the level $156$i can be populated, ionization will happen. In order to calculate the population on it transferred from the initial level $|\mathsf{n}_1s\rangle$~(or $|\mathsf{n}_2s\rangle$) of the state $|1\rangle$, we may then simulate the time dynamics starting from the state $\sin\beta|\mathsf{n}_1s\rangle$, with the results given in Fig.~\ref{fig005}. Obviously, almost no population ever goes beyond the transition $k=2$. In fact, only the level $62p$ is populated by a one-photon transition, which means that ionization does not happen in our model. Similar results hold for the transition chain from $|\mathsf{n}_2s\rangle$.

The ionization during the gate sequence should have even weaker effect. For example, if only microwave field \circled{1} is present, $\delta_{2}^{(i)}$ of Table~\ref{table1} will only increase because only the field \circled{1} is available to build up a similar transition chain. This gives an even smaller population on the topmost level that leads to ionization.

Notice that the result in Fig.~\ref{fig005} does not show what happens in reality, since the level $62p$ gives a two-photon resonance with an effective Rabi frequency $\Omega^{(2)}$ which was studied in the last section. In other words, the level $62p$ should not be populated as much as in Fig.~\ref{fig005}.

\begin{figure}
\includegraphics[width=3.1in]
{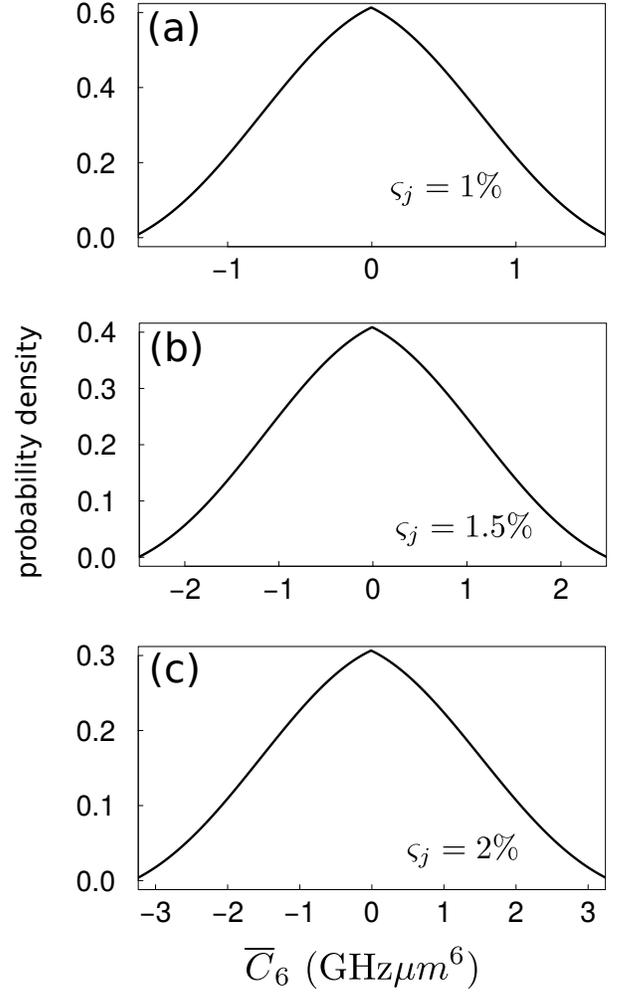}
\caption{Probability distribution of $\overline{C}_6$  when the dressing microwave Rabi frequency $\Omega_j'$ obeys a Gaussian distribution with a standard deviation $\varsigma_j\Omega_j$, where $j=1$ or $2$. $\varsigma_j$ is $1\%, 1.5\%$ and $2\%$ in (a), (b) and (c), respectively. \label{figC6}}
\end{figure}

\section{Influence of amplitude stability of the dressing microwave fields upon the dark state}\label{App_K}
The Rabi frequencies $\Omega_{1(2)}$ that define $|1\rangle $ and hence the dark state $|11\rangle$, are proportional to the amplitudes of the dressing microwave fields \circled{1} and \circled{2}. Amplitude fluctuations cause fluctuations of $\beta$ around $\beta_0$, and result in a nonzero value of $\overline{C}_6$ and vdWI for the state $|11\rangle$. This will lead to an incomplete population of $|11\rangle$ during initialization due to the blockade effect. The gate error due to this effect will be discussed in this section.

Consider a case where the amplitudes of the two microwave fields obey a Gaussian distribution, so that the distribution of the Rabi frequencies $\Omega_1'$ and $\Omega_2'$ is given by 
\begin{eqnarray}
f_j(\Omega_j') &=& \text{exp}[-(\Omega_j'-\Omega_j)^2/(2\varsigma_j^2\Omega_j^2)]/\sqrt{2\varsigma_j^2\Omega_j^2\pi},\nonumber\\&&~ j\in\{1,2\}.\nonumber
\end{eqnarray}
Here we assume microwave amplitude fluctuations with a standard deviation around $10^{-2}$, i.e., we choose $\varsigma_j=1\%$. The probability density for $(\Omega_1',\Omega_2')$ is thus $f_1(\Omega_1')f_2(\Omega_2')$, where the corresponding $\overline{C}_6$ is given by
  \begin{eqnarray}
\overline{C}_6(\beta')= a\sin^4\beta'+ d\cos^4\beta'+2(b+c)\sin^2\beta'\cos^2\beta',\nonumber
  \end{eqnarray}
  and $\tan\beta' = \Omega_2'/\Omega_1'$. With these settings, the probability distribution of $\overline{C}_6(\beta')$ is shown in Fig.~\ref{figC6}(a), which tells us that most $\overline{C}_6(\beta')$ falls inside the interval $\{-1,1\}$~GHz~$\mu m^6$. With $L=4.4\mu m$, this means that the residual vdWI between the two atoms for the state $|11\rangle$ is usually smaller than $V_{\text{m}}=0.14$~MHz. If the initialization process happens with a laser Rabi frequency $\kappa=10$~MHz, the rotation error characterizing the failure of preparing the state $|11\rangle$ is $4(V_{\text{m}}/\kappa)^2$~\cite{Saffman2005}, which contributes to the gate error by $4(V_{\text{m}}/\kappa)^2/4\approx2\times10^{-4}$, much smaller than other errors in the gate protocol. Another consequence of the microwave field amplitude fluctuations is that the state $|1\rangle$ deviates from the designed one: $|\langle\mathsf{n}_js |1\rangle|^2$ can have variance of around $\varsigma_j^2$. With $\varsigma_j=1\%$, we conclude that such errors are also negligible.

When the amplitude fluctuation increases, the probability distribution of $\overline{C}_6(\beta')$ becomes wider, as shown in Fig.~\ref{figC6}(b) and~(c). One can find that for $\varsigma_j=2\%$ as shown in Fig.~\ref{figC6}(c), most $\overline{C}_6(\beta')$ falls inside the interval $\{-2,2\}$~GHz~$\mu m^6$. With $L=4.4\mu m$, this means that most of the residual vdWI is smaller than $V_{\text{m}}=0.28$~MHz. As a result, the gate error from not preparing the state $|11\rangle$ is about $7.6\times10^{-4}$. Meanwhile, $|\langle\mathsf{n}_js |1\rangle|^2$ can have a variance of about $4\times10^{-4}$. This means that the gate error will increase by about $10^{-3}$, comparable to the intrinsic gate fidelity error in the order of $10^{-3}$. However, if $\varsigma_j<2\%$, the extra gate fidelity error due to the amplitude fluctuation of the microwave fields is mostly smaller than $10^{-3}$ and can be neglected.  

  \section{Fluctuation of atomic positions}\label{App_L}
  The gate will suffer from a spatial distribution of the atoms in their respective traps because the phase shift accumulated during the wait period depends on the blockade shift $\mathbb{B}$ of the state $|22\rangle$. Here we will analyze the error due to this effect, where the atomic spatial distribution is determined by the trap parameters and the atomic motional states.

\begin{figure}
\includegraphics[width=2.1in]
{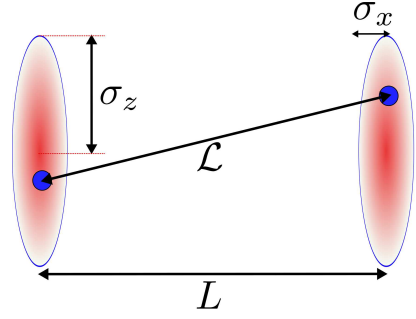}
\caption{Schematic of position fluctuation of atoms. An atom is mainly trapped inside the ellipse. $\sigma_j^2$ is the variance of the atomic position along the $j$-axis, where $j=x,y$ or $z$. Here $\sigma_x=\sigma_y$. \label{fig-position01}}
\end{figure}

A commonly used method of trapping a single neutral atom in Rydberg experiments is far-off-resonance optical trap~\cite{Isenhower2010}, or optical tweezer~\cite{Wilk2010}. With optical tweezers, the authors of Ref.~\cite{Kaufman2012} successfully employed Raman sideband cooling to cool a $^{87}$Rb atom to its motional ground state. The parameters characterizing an optical trap include the trap depth $U$ and the $1/e^2$ beam radius $w$, which further determines the qubit's oscillation frequencies $\{\omega_x,\omega_y,\omega_z\}$, and the averaged variances, $\{\sigma_x^2,\sigma_y^2,\sigma_z^2\}$, of its position.

When the motional state of a trapped neutral atom is thermal, i.e., $k_BT_a/2\geq \hbar \omega_j$, $j=x,y,z$, the position fluctuation of the trapped atom can be as large as several $\mu$m~\cite{Isenhower2010}. Because the atomic separation $L$ is only several $\mu$m in our gate, we conclude that in this regime the gate protocol will have sizable errors due to the position fluctuations of atoms.

When the motional state of a trapped neutral atom is cooled to the point that $k_BT_a/2\leq \hbar \omega_j$, $j=x,y,z$, the trapped atom will be in its motional ground state characterized with zero vibrational excitation~\cite{Kaufman2012}. For motion in the $x$ direction, the atomic eigenfunctions of motion are
\begin{eqnarray}
\psi_0(x) &=& \frac{\sqrt\alpha}{\pi^{1/4}}e^{-\alpha^2 x^2/2}, \nonumber\\
\psi_1(x) &=&\alpha x \frac{\sqrt{2\alpha}}{\pi^{1/4}}e^{-\alpha^2 x^2/2}, \nonumber\\
\cdots,\nonumber
\end{eqnarray}
where $\alpha =\sqrt\frac{\mu\omega_j}{\hbar} $, $j=x,y$ or $z$, and $\mu$ is the mass of the atom, (equal to $87u$, with $u$ the atomic mass unit, for a $^{87}$Rb atom). The corresponding position variances are
\begin{eqnarray}
\sigma_j^2 &=&\left(n+\frac{1}{2}\right) \frac{\hbar}{\mu\omega_j}.\nonumber
\end{eqnarray}
Below we consider the case of $n=0$.

For a typical frequency $\omega_j=150\times 2\pi$~kHz, we have $\sigma_j=19.6$~nm. Similarly for the other two directions of $y$ and $z$. As can be easily verified, the change of the two-atom orientation away from the quantization axis is minor with such $\sigma_j$, thus the C-six coefficient of the dark state is still zero. Because the dipole-dipole interaction is also much smaller than the microwave field induced energy gaps in Eq.~(\ref{eqA10}), the state $|11\rangle$ remains a good dark state with such magnitude of $\sigma_j$. Thus the position fluctuations of the qubits do not influence the establishment of the dark state.

Consider an optical tweezer studied in Ref.~\cite{Kaufman2012} for trapping and cooling an $^{87}$Rb atom, the trap frequencies along the $x$ and $y$ directions are theoretically given by
\begin{eqnarray}
\omega_x\approx\omega_y = \frac{2}{w}\sqrt{U/\mu},\nonumber
\end{eqnarray}
which gives $\omega_x\approx\omega_y=2\pi\times153$~kHz if $U=1.4$mK and $w=0.76\mu$m, as from the example of Ref.~\cite{Kaufman2012}. The measured result in Ref.~\cite{Kaufman2012} is $\{\omega_z,\omega_x,\omega_y\}=$$\{30,154,150\}$~kHz, which means that $\omega_z\approx \omega_x/5$. Concerning whether it is possible to trap two neutral atoms as close as the $L$'s chosen in this work, we note that  the authors in Ref.~\cite{Lester2015} created arrays of deeper optical tweezers, where each trap is characterized by $U=73$~MHz$\sim3.5$~mK and $w=0.71\mu$m, and successfully loaded $^{87}$Rb atoms in small arrays with optical lattice constant as small as $1.7\mu$m.

As shown in Fig.~\ref{fig-position01}, the actual distance $\mathcal{L}$ between the two atoms can be different from $L$. To describe this, we denote the locations of the trap centers for the control and target qubits as $(0,0,0)$ and $(L,0,0)$, respectively. When the vibrational state of the control qubit inside an optical tweezer is the ground state, the distribution of its actual location is $|\psi_x\psi_y\psi_z|^2$, i.e.,
\begin{eqnarray}
  f_c(x_c,y_c,z_c) &=& \text{exp} \left[-0.5(x_c^2+y_c^2)/\sigma_x^2 -0.5z_c^2/\sigma_z^2\right]\nonumber\\
  &&/\left[\sigma_x^2\sigma_z(2\pi)^{3/2}\right],\nonumber
\end{eqnarray}
while that for the target is
\begin{eqnarray}
  f_t(x_t,y_t,z_t) &=& \text{exp} \left[-0.5((x_t-L)^2+y_t^2)/\sigma_x^2 -0.5z_t^2/\sigma_z^2\right]\nonumber\\
  &&/\left[\sigma_x^2\sigma_z(2\pi)^{3/2}\right].\nonumber
\end{eqnarray}
Here variables with the subscript c~(t) denote those for the control~(target) qubit.

\begin{figure}
\includegraphics[width=3.3in]
{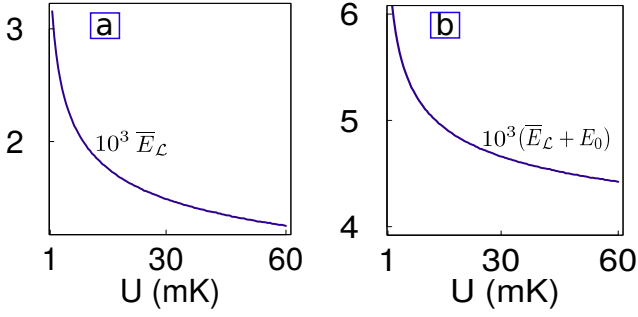}
\caption{(a) Scaled gate fidelity error due to the fluctuation of atomic positions as a function of the trap depth when $L=5.2\mu m$. (b) Total gate fidelity error $E_0+\overline{E}_{\mathcal{L}}$ scaled by $10^3$ when $L=5.2\mu m$, where $E_0=3.17\times10^{-3}$.  \label{p2}}
\end{figure}

For different runs of the gate cycles, the fluctuation of the atomic location will give different $\mathcal{L}$'s and different orientations of the two-atom axis relative to the quantization axis $x$. For $\sigma_z\ll L$, we mainly focus on $\mathcal{L}$'s fluctuation and its consequence upon the gate performance. The blockade shift of the state $|22\rangle$, where $|2\rangle=|\mathsf{n}_2s\rangle$, is almost isotropic along the whole solid angle~\cite{Walker2008}, thus $\mathbb{B}(\mathcal{L})=C_6/\mathcal{L}^6$, where $C_6=113$~GHz~$\mu \text{m}^6$ as from the paragraph following Eq.~(\ref{eq03}) of the main text. For a $C_Z$ gate, the deviation of the blockade shift from $\mathbb{B}(L)$ will contribute an extra error $E_{\mathcal{L}}(\mathcal{L})$ to the total gate fidelity error, where
\begin{eqnarray}
  E_{\mathcal{L}}(\mathcal{L})&=&|\text{exp}\left[-i \mathbb{B}(\mathcal{L}) T \right]-\text{exp}\left[-i \mathbb{B}(L) T \right]|/4.\nonumber
\end{eqnarray}
Notice that $\mathcal{L}^2=(x_c-x_t)^2+(y_c-y_t)^2+(z_c-z_t)^2$. The average $E_{\mathcal{L}}(\mathcal{L})$ is
\begin{eqnarray}
  \overline{E}_{\mathcal{L}}&=&\int dx_c\int dy_c \cdots \int dy_t\int dz_t E_{\mathcal{L}}(\mathcal{L})f_c f_t .  \nonumber
\end{eqnarray}
The numerical integration above as a function of $U$ can be performed by Monte Carlo integration, where a test can be made by checking if $\overline{E}_{\mathcal{L}}$ becomes unit when we set $E_{\mathcal{L}}(\mathcal{L})=1$ in the integral above. The numerical result of $\overline{E}_{\mathcal{L}}$ is presented in Fig.~\ref{fig004}(b) of the main text, where $L=4.4\mu m$ is used. From Fig.~\ref{fig004}(b), one can find that $\overline{E}_{\mathcal{L}}$ drops from $3.0\times10^{-3}$ to $1.5\times10^{-3}$ when $U=3.5$~mK, as realized in~\cite{Lester2015}, increases to $60$~mK. Even with a very small depth of $U=1.4$~mK as in~\cite{Kaufman2012}, we can still have $\overline{E}_{\mathcal{L}}=3.7\times10^{-3}$. Also, it should be noticed that for a larger $L$, a smaller $\overline{E}_{\mathcal{L}}$ should be expected. For instance, consider $L=5.2\mu$m, where an optimized gate error $E_0=3.17\times10^{-3}$ with a gate operation time $94$~ns was found, with $\overline{E}_{\mathcal{L}}$ ignored. With this $L$, we have $\overline{E}_{\mathcal{L}}\in\{3.16,1.25\}\times10^{-3}$ when $U\in\{1.5,60\}$~mK, shown in Fig.~\ref{p2}(a). The total~(actual) gate fidelity error $\overline{E}_{\mathcal{L}}+E_0$ can be $4.4\times 10^{-3}$ when $U=60$~mK for this $L$, as shown in Fig.~\ref{p2}(b).

Setting the trap depth $U$ equal to a few times of $10$~mK is feasible for an optical trap~\cite{Saffman2016}. Moreover, the necessary conditions to (i) cool neutral atoms to their motional ground states~\cite{Kaufman2012} and (ii) load neutral atoms efficiently to optical tweezer lattice of a small lattice constant~\cite{Lester2015} were already demonstrated in experiments. So, it is possible to realize a fast and accurate two-qubit $C_Z$ quantum gate with our protocol.



\end{document}